\documentclass[fleqn,usenatbib]{mnras}

\usepackage{newtxtext,newtxmath}

\usepackage[T1]{fontenc}

\DeclareRobustCommand{\VAN}[3]{#2}
\let\VANthebibliography\thebibliography
\def\thebibliography{\DeclareRobustCommand{\VAN}[3]{##3}\VANthebibliography}

\usepackage{graphicx}	
\usepackage{amsmath}	
\usepackage[flushleft]{threeparttable}
\usepackage{caption}
\usepackage{longtable}
\usepackage{makecell}
\usepackage{lineno}
\usepackage[version=4]{mhchem}
\usepackage{amsmath}

\title[Amides in G+0.693]{Amides inventory towards the G+0.693-0.027 molecular cloud}

\author[S. Zeng et al.]{
S. Zeng,$^{1}$\thanks{E-mail: shaoshan.zeng@riken.jp}
V. M. Rivilla,$^{2}$
I. Jim\'enez-Serra,$^{2}$
L. Colzi,$^{2}$
J. Mart\'in-Pintado,$^{2}$
\newauthor{
B, Tercero,$^{3,4}$
P. de Vicente,$^{4}$
S. Mart\'in,$^{5,6}$
and M. A. Requena-Torres$^{7}$
}
\\
$^{1}$Star and Planet Formation Laboratory, Cluster for Pioneering Research, RIKEN, 2-1 Hirosawa, Wako, Saitama, 351-0198, Japan\\
$^{2}$Centro de Astrobiolog\'ia (CSIC-INTA), Ctra. de Ajalvir Km. 4, 28850, Torrej\'on de Ardoz, Madrid, Spain \\
$^{3}$Observatorio Astron\'omico Nacional (OAN-IGN), Calle Alfonso XII, 3, 28014 Madrid, Spain\\
$^{4}$Observatorio de Yebes (OY-IGN), Cerro de la Palera s/n, Yebes, 19141, Guadalajara, Spain\\
$^{5}$European Southern Observatory, Alonso de C\'ordova 3107, Vitacura 763-0355, Santiago, Chile\\
$^{6}$Joint ALMA Observatory, Alonso de C\'ordova 3107, Vitacura 763-0355, Santiago, Chile\\
$^{7}$Department of Physics, Astronomy and Geosciences, Towson University, Towson, MD 21252, USA
}

\date{Accepted XXX. Received YYY; in original form ZZZ}

\pubyear{2023}

\begin{document}
\label{firstpage}
\pagerange{\pageref{firstpage}--\pageref{lastpage}}
\maketitle

\begin{abstract}
Interstellar amides have attracted significant attentions as they are potential precursors for a wide variety of organics essential to life. However, our current understanding of their formation in space is heavily based on observations in star-forming regions and hence the chemical networks lack the constraints on their early origin. In this work, unbiased sensitive spectral surveys with IRAM 30m and Yebes 40m telescopes are used to systematically study a number of amides towards a quiescent Galactic Centre molecular cloud, G+0.693-0.027. We report the first detection of acetamide (CH$_3$C(O)NH$_2$) and trans-N-methylformamide (CH$_3$NHCHO) towards this cloud. In addition, with the wider frequency coverage of the survey, we revisited the detection of formamide (NH$_2$CHO) and urea (carbamide; NH$_2$C(O)NH$_2$), which had been reported previously towards G+0.693-0.027. Our results are compared with those present in the literature including recent laboratory experiments and chemical models. We find constant abundance ratios independently of the evolutionary stages, suggesting that amides related chemistry is triggered in early evolutionary stages of molecular cloud and remain unaffected by the warm-up phase during the star formation process. Although a correlation between more complex amides and NH$_2$CHO have been suggested, alternative formation routes involving other precursors such as acetaldehyde (CH$_3$CHO), methyl isocyanate (CH$_3$NCO) and methylamine (CH$_3$NH$_2$) may also contribute to the production of amides. Observations of amides together with these species towards a larger sample of sources can help to constrain the amide chemistry in the interstellar medium. 
\end{abstract}



\section{Introduction}
Amides are chemical species that contain a carbonyl group linked to a nitrogen atom (R$_1-$C(=O)$-$N$- \rm R^{\prime\prime}$), which can also be referred to as peptide bonds during the formation of proteins, where a large number of amino acids bind with each other through peptide chains. Amides are thus among the essential components in the fundamental building blocks of life that make up the structural components of living cells and regulate biochemical processes \citep[][]{Ruiz-Mirazo2014}. The presence of amino acids on the prebiotic Earth is widely accepted, either coming from endogenous chemical processes \citep[e.g.][]{Patel2015} or forming in the interstellar medium (ISM), and with their subsequent delivery to Earth \citep[e.g.][]{Altwegg2016}. As such, understanding the formation of amides in the ISM can be relevant for studying the chemical origin of life. 

Formamide (NH$_2$CHO) was the first amide detected in the ISM \citep[towards Sgr B2;][]{Rubin1971} and has then been robustly detected in several star-forming regions \citep[see e.g. very recent observations by][and see review by \citealt{Lopez-Sepulcre2019} for a detailed list of astronomical observations]{Colzi2021,Ligterink2020, Ligterink2022}. Ever since the detection of formamide, amides with increasing complexity such as acetamide \citep[CH$_3$C(O)NH$_2$;][]{Hollis2006,Halfen2011,Belloche2017,Ligterink2020,Ligterink2022,Colzi2021}, and N-methylformamide \citep[CH$_3$NHCHO;][]{Belloche2017,Ligterink2020,Ligterink2022,Colzi2021} have been detected in the ISM, although mostly exclusively towards high-mass star-forming regions. The detection of another amide, urea (aka carbamide, NH$_2$C(O)NH$_2$), has also been confirmed towards the Galactic Centre (GC) source Sgr B2(N1) \citep{Belloche2019} and the molecular cloud G+0.693-0.027 \citep{Jimenez-Serra2020}. However, searches of more complex amides such as cyanoformamide (NH$_2$COCN) and glycolamide (HOCH$_2$C(O)NH$_2$) in the ISM have been attempted without success \citep[][]{Colzi2021,Ligterink2022,Sanz-Novo2022}.

To get insight into how complex the interstellar chemistry is and what prebiotic molecules can be synthesised in space, it is essential to understand the chemical processes that result in the formation of simpler amides detected in the ISM. For the aforementioned amides, both gas-phase and grain formation routes have been discussed to explain their observed abundance in the regions where they reside \citep[e.g.][]{ Jones2011,Skouteris2017,Quenard2018,Belloche2019,Douglas2022,Garrod2022}. 

For example, tight column density correlation is found between NH$_2$CHO and CH$_3$C(O)NH$_2$, which might indicate these two species are formed through related chemical processes or have the same responses to similar physical conditions \citep{Colzi2021,Ligterink2022}. However, detection of several amides from the same regions are still relatively sparse and they are biased towards the physical conditions of star-forming regions. The chemical networks that include amides may therefore lack of further constraints to elucidate the chemical link between these species and the interplay with the physical conditions especially at earliest stage of star formation. 

In this work, the most complete inventory of amide species is investigated in detail towards the quiescent molecular cloud G+0.693-0.027 (hereafter G+0.693). G+0.693 is located within the Sgr B2 star-forming complex in the center of our Galaxy. Despite the fact it does not shown signs of on-going star-formation activity \citep[][]{Zeng2020}, it represents one of the most chemically rich sources in our Galaxy. With plentiful of nitrogen-bearing species detected towards G+0.693, the family of nitrile ($-$CN) and amines ($-$NH$_2$) as well as several complex organic molecules (COMs) that are of pre-biotic interest have also been reported in previous studies \citep{Zeng2018,Rivilla2019,Rivilla2020,Rivilla2021a,Rivilla2021b,Rivilla2022c,Bizzocchi2020,Rodriguez-Almeida2021b,Zeng2021}. Following up the detection of NH$_2$CHO \citep{Zeng2018} and NH$_2$C(O)NH$_2$ \citep{Jimenez-Serra2020}, additional transitions of these species along with the first detection of CH$_3$C(O)NH$_2$ and CH$_3$NHCHO towards G+0.693 is presented in Section \ref{observations} and \ref{results}. The possible formation routes for each amide are discussed in \ref{discussion} whilst the conclusions are given in Section \ref{conclusions}.

\begin{table*}
   \centering
    \caption[]{Physical parameters obtained from the best LTE fit in \textsc{madcuba} and derived molecular abundance with respect to H$_2$.}
        \label{tab:mol_parameters}
    \begin{threeparttable}
    \begin{tabular}{cccccc}
         \hline
         \hline
         Molecule & $T\rm_{ex}$ & $\rm \nu_{LSR}$ & FWHM  & $\textit{N}_{\rm tot}$ & $X^{a}$ \\
         & (K) & (km s$^{-1}$) & (km s$^{-1}$) & ($\times$10$^{13}$ cm $^{-2}$) &  ($\times$10$^{-10}$)\\
         \hline
         NH$_2$$^{13}$CHO & 5.0$\pm$0.3 & 69.1$\pm$0.6 & 21$\pm$1 & 0.62$\pm$0.04 & 0.46$\pm$0.03 \\
         NH$_2$CHO$^c$ & - & - & - & 25$\pm$1 & 18$\pm$1 \\
         \hline
         CH$_3$C(O)NH$_2$ A & 7.4$\pm$0.2 & 68.7$\pm$0.3 & 19.5$\pm$0.6 & 3.4$\pm$0.1 & 2.50$\pm$0.07 \\
         CH$_3$C(O)NH$_2$ E & 7.8$\pm$0.1 & 68.7$\pm$0.2 & 21.6$\pm$0.4 & 8.1$\pm$0.1 & 6.0$\pm$0.1 \\
         CH$_3$C(O)NH$_2$ (A+E) & - & - & - &  11.5$\pm$0.2 & 8.5$\pm$0.1 \\
         \hline
         trans-CH$_3$NHCHO & 7.1$\pm$0.4 & 68.2$\pm$0.5 & 19$\pm$1 & 4.3$\pm$0.4 & 3.2$\pm$0.3 \\
         \hline
         NH$_2$C(O)NH$_2$ & 8.0$\pm$0.7 & 69$^d$ & 20$^b$ & 0.71$\pm$0.05 & 0.52$\pm$0.05 \\
         \hline
    \end{tabular}
    \begin{tablenotes}
            \item[a] $N_{\rm H_2}$=1.35$\times$10$^{23}$ cm$^{-2}$ \citep{Martin2008}.
            \item[b] Value of FWHM fixed in the \textsc{madcuba} fit.
            \item[c] Column density and abundance derived from the $^{13}$C-isotopologue by multiplying $^{12}$C/$^{13}$C=40 (Colzi et al. in prep).
            \item[d] Value of $\rm \nu_{LSR}$ fixed in the \textsc{madcuba} fit.
        \end{tablenotes}
     \end{threeparttable}
   \end{table*}

\section{Observations}
\label{observations}
The high-sensitivity spectral surveys towards G+0.693 molecular cloud were carried out with IRAM 30$\,$m\footnote{IRAM is supported by INSU/CNRS (France), MPG (Germany), and IGN (Spain)} and Yebes 40$\,$m\footnote{The 40$\,$m radiotelescope at Yebes Observatory is operated by the Spanish Geographic Institute (IGN, Ministerio de Transportes, Movilidad y Agenda Urbana.) \url{http://rt40m.oan.es/rt40m en.php}} telescopes. The IRAM 30$\,$m observations were performed during three sessions in 2019: 10$\rm ^{th}-$16$\rm ^{th}$ of April, 13$\rm ^{th}-$19$\rm ^{th}$ of August, and 11$\rm ^{th}-$15$\rm ^{th}$ of December whilst the observations with the Yebes 40$\,$m telescope were carried out through project 20A008 between 3$\rm ^{rd}-$9$\rm ^{th}$ and 15$\rm ^{th}-$22$\rm ^{nd}$ of February 2020. The observations were centred at $\alpha$(J2000) = 17$\rm^h$47$\rm^m$22$\rm^s$, $\delta$(J2000) = -28$\rm^{\circ}$21$\rm^{\prime}$27$\rm^{\prime\prime}$. The position switching mode was used in all observations with the reference position located at $\Delta\alpha$, $\Delta\delta$ = $-$885$^{\prime\prime}$,290$^{\prime\prime}$ with respect to the source position. The IRAM 30$\,$m observations covered the frequency ranges of 71.8$-$116.7$\,$GHz, 124.8$-$175.5 GHz, 199.8$-$222.3$\,$GHz, and 223.3$-$238.3$\,$GHz. The half$-$power beam width (HPBW) of the telescope spanned between $\sim$34$^{\prime\prime}-11^{\prime\prime}$. The spectral coverage of Yebes 40$\,$m observations ranged from 31.3$\,$GHz to 50.6$\,$GHz. The HPBW of the telescope was in a range of $\sim$54$^{\prime\prime}-36^{\prime\prime}$. The data from both surveys were smoothed to a velocity resolution of $\sim$1.0$-$2.6$\,$km s$^{-1}$ and the intensity was measured in the unit of antenna temperature, T$_{\rm A}^*$, as the molecular emission towards G+0.693 is extended over the beam \citep{Requena-Torres2006,Requena-Torres2008,Zeng2020}. Depending on the frequency, the noise of the spectra for the Yebes 40$\,$m data is 1.0$\,$mK, whilst in some intervals can increase up to 4.0$-$5.0$\,$mK; whereas 1.3 to 2.8$\,$mK (71$-$90$\,$GHz), 1.5 to 5.8$\,$mK (90$-$115$\,$GHz), $\sim$10 mK (115$-$116$\,$GHz), 3.1 to 6.8$\,$mK (124$-$175$\,$GHz), and 4.5 to 10.6$\,$mK (199$-$238$\,$GHz), for the IRAM 30$\,$m data. More detailed information of these observations is provided in \citet{Zeng2020,Rodriguez-Almeida2021a}.

   \begin{figure*}
   \centering
  \includegraphics[width=\textwidth]{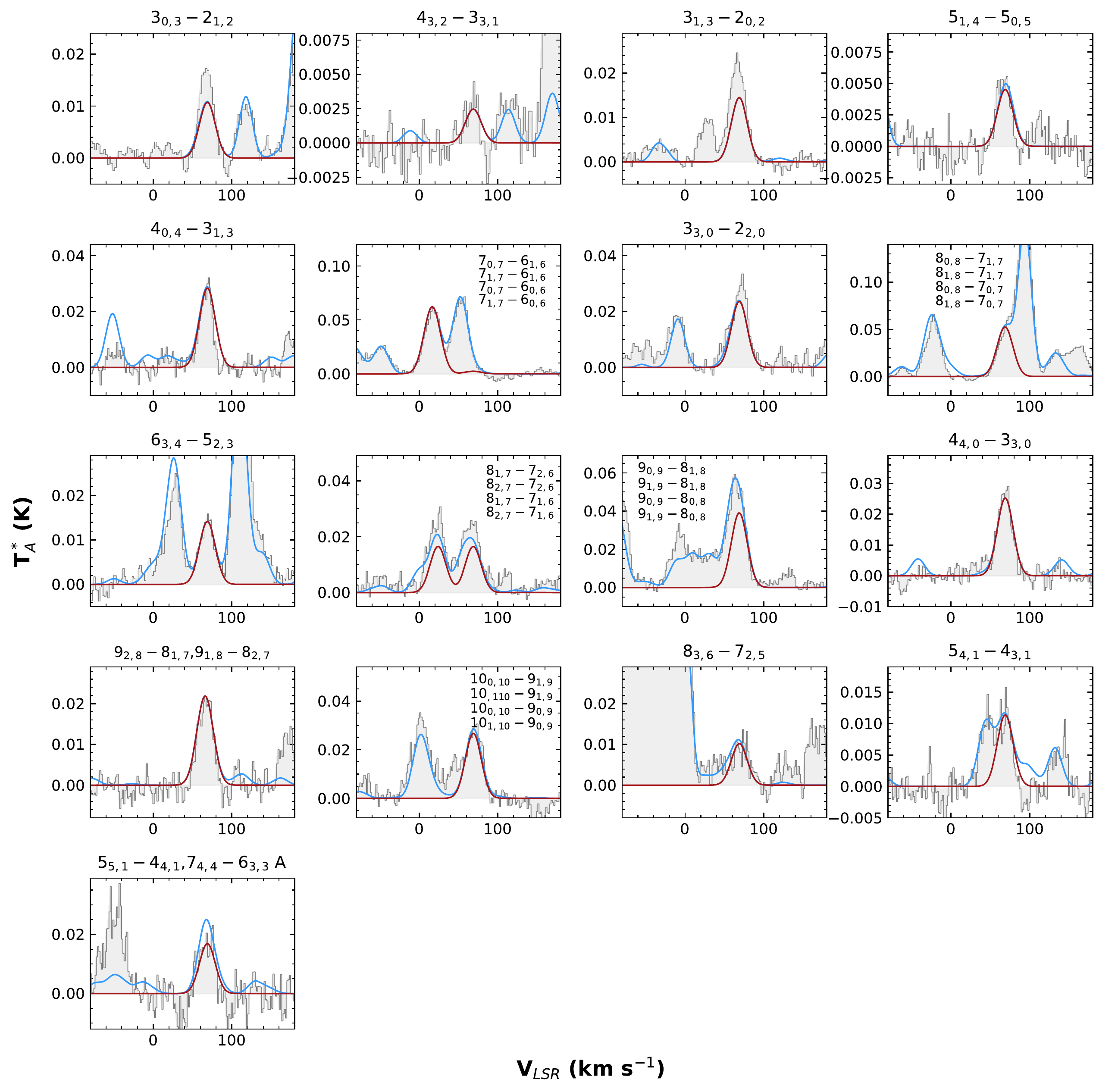}
   \caption{Detected E-state transitions of CH$_3$C(O)NH$_2$ towards G+0.693. The observed spectrum is plotted in grey, with the best LTE fit obtained with MADCUBA overplotted in red and the synthetic spectrum considering contribution from all the species identified in the source is indicated by the blue line. The quantum number of each transition is given on the top of each panel.}
              \label{fig:CH3CONH2_E_spectra}%
    \end{figure*}
    
   \begin{figure*}
   \centering
  \includegraphics[width=\textwidth]{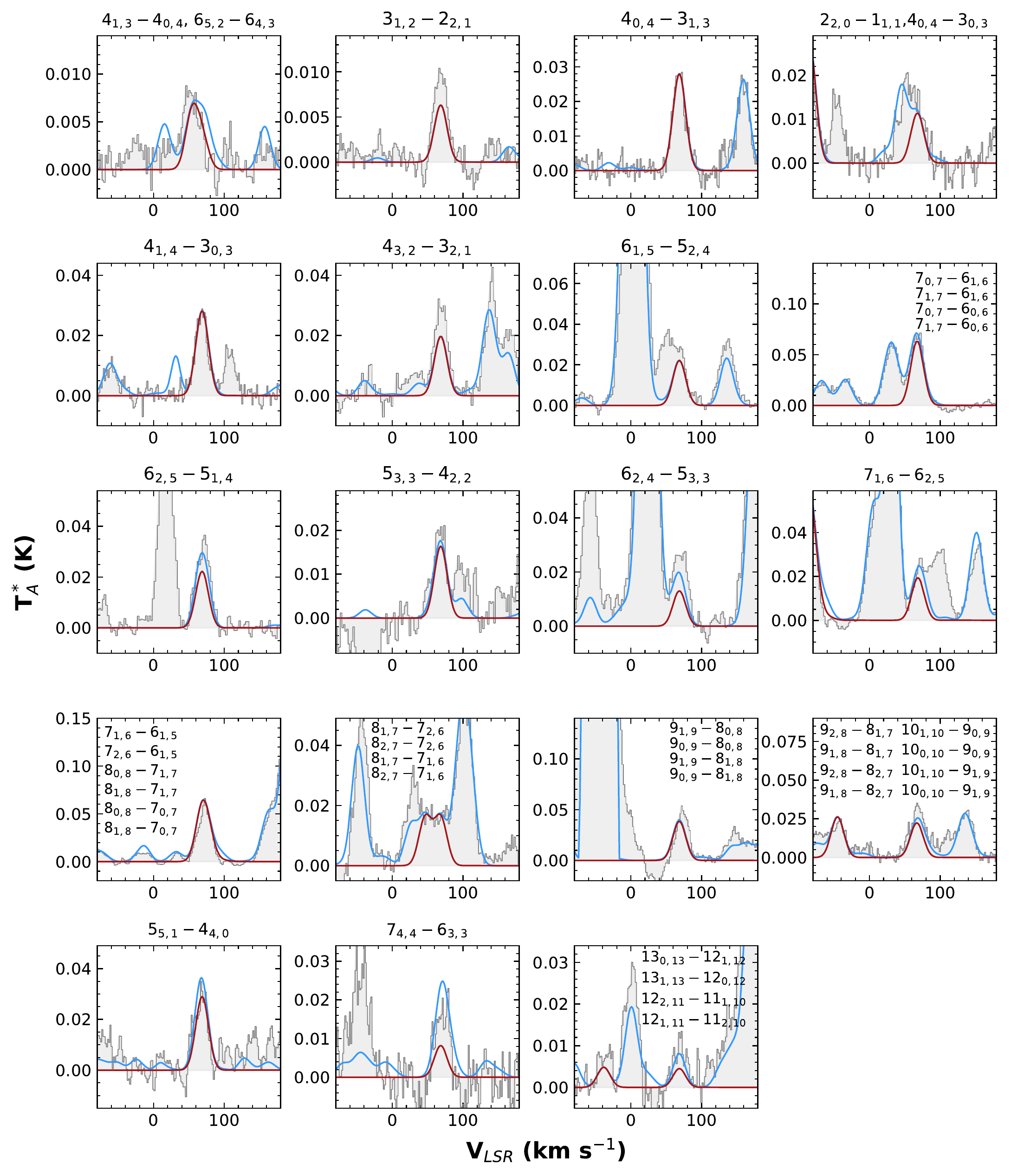}
   \caption{Detected A-state transitions of CH$_3$C(O)NH$_2$ towards G+0.693. The observed spectrum is plotted in grey, with the best LTE fit obtained with MADCUBA overplotted in red and the synthetic spectrum considering contribution from all the species identified in the source is indicated by the blue line. The quantum number of each transition is given on the top of each panel.}
              \label{fig:CH3CONH2_A_spectra}%
    \end{figure*}

   \begin{figure*}
   \centering
  \includegraphics[width=\textwidth]{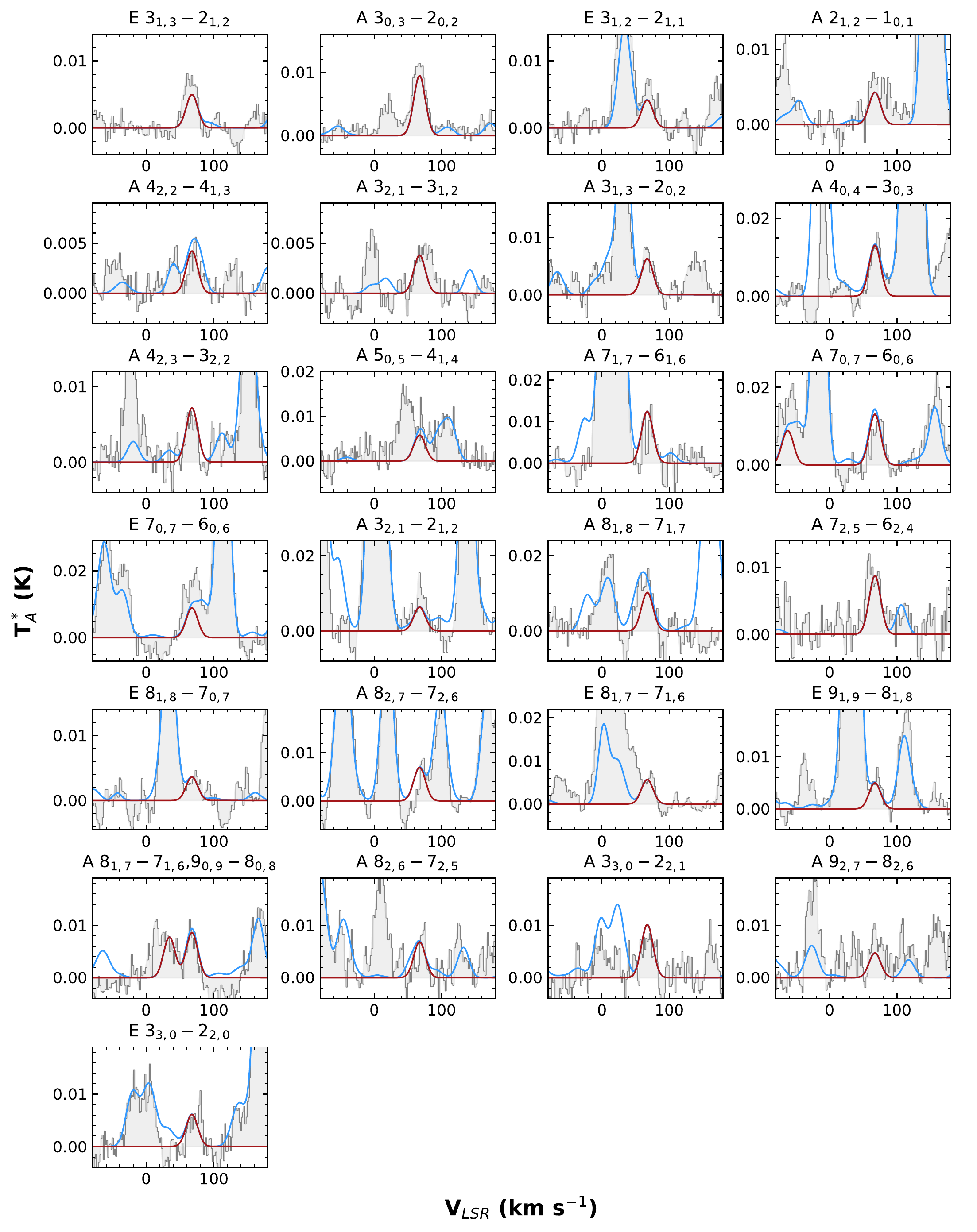}
   \caption{Detected transitions of CH$_3$NHCHO towards G+0.693. The observed spectrum is plotted in grey, with the best LTE fit obtained with MADCUBA overplotted in red and the synthetic spectrum considering contribution from all the species identified in the source is indicated by the blue line. The quantum number of each transition is given on the top of each panel.}
              \label{fig:CH3NHCHO_spectra}%
    \end{figure*}

\section{Analysis and results}
\label{results}
\begin{figure*}
   \centering
  \includegraphics[width=\textwidth]{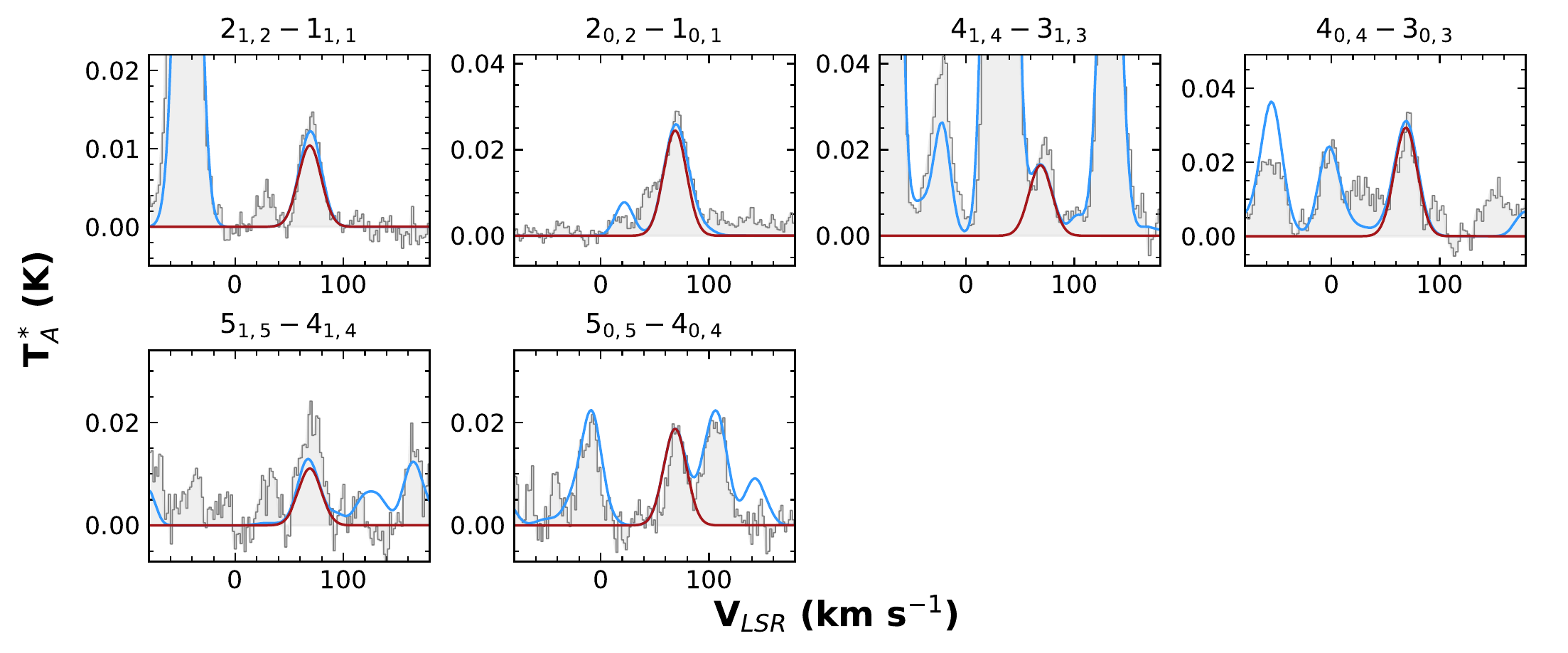}
   \caption{Detected transitions of NH$_2^{13}$CHO towards G+0.693. The observed spectrum is plotted in grey, with the best LTE fit obtained with MADCUBA overplotted in red and the synthetic spectrum considering contribution from all the species identified in the source is indicated by the blue line. The quantum number and the upper state energy of each transition is given on the top of each panel.}
              \label{fig:NH213CHO_spectra}%
    \end{figure*}

\begin{figure*}
   \centering
  \includegraphics[width=\textwidth]{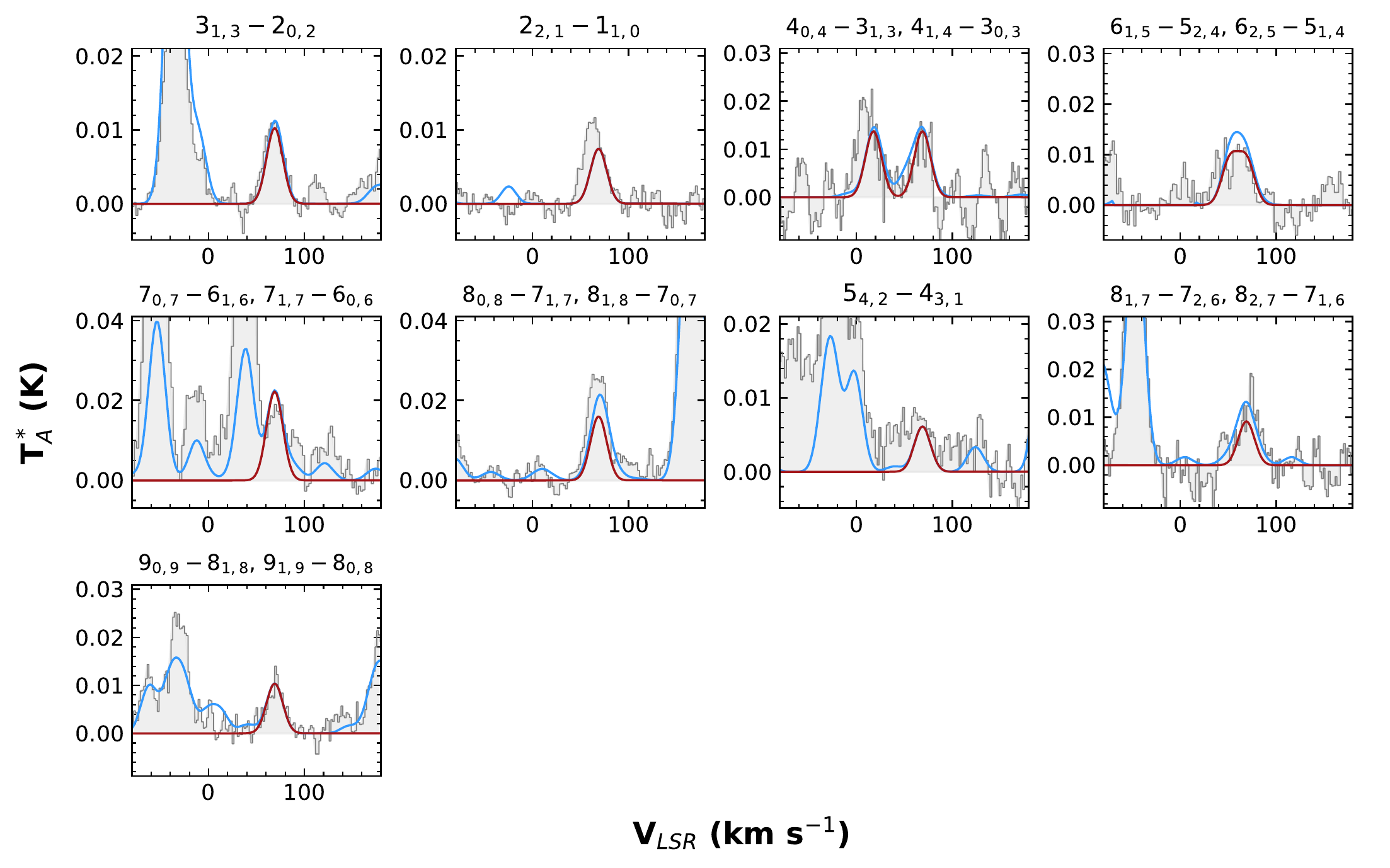}
   \caption{Detected transitions of NH$_2$C(O)NH$_2$ towards G+0.693. The observed spectrum is plotted in grey, with the best LTE fit obtained with MADCUBA overplotted in red and the synthetic spectrum considering contribution from all the species identified in the source is indicated by the blue line. The quantum number of each transition is given on the top of each panel.}
              \label{fig:H2NCONH2_spectra}%
    \end{figure*}

The line identification and the analysis were carried out using the SLIM (Spectral Line Identification and Modelling) tool implemented within the \textsc{madcuba} package\footnote{Madrid Data Cube Analysis on ImageJ is a software developed at the Center of Astrobiology (CAB) in Madrid; \url{http://cab.inta-csic.es/madcuba/}}\citep[version 10/03/2022,][]{Martin2019}. SLIM uses the spectroscopic entries from the Cologne Database for Molecular Spectroscopy\footnote{\url{https://cdms.astro.uni-koeln.de/classic/}} \citep[CDMS,][]{Endres2016}, the Jet Propulsion Laboratory\footnote{\url{https://spec.jpl.nasa.gov/ftp/pub/catalog/catdir.html}} \citep[JPL,][]{Pickett1998}, and our own database with added entries for species that are not included in the previous catalogue by using available spectroscopic literature. SLIM generates synthetic spectra under the assumption of Local Thermodynamic Equilibrium (LTE) conditions and taking into account the line opacity. For the molecules presented in this work, the spectroscopic information is given in Table \ref{tab:mol_spec} in Appendix A. In order to properly evaluate the potential line contamination, the emission from over 125 molecules identified towards G+0.693 have been considered \citep[][]{Requena-Torres2006,Requena-Torres2008,Rivilla2018,Rivilla2019,Rivilla2020,Rivilla2021a,Rivilla2021b,Rivilla2022b,Rivilla2022c,Rivilla2022a,Bizzocchi2020,Jimenez-Serra2020,Jimenez-Serra2022,Rodriguez-Almeida2021b,Rodriguez-Almeida2021a,Zeng2018,Zeng2021}. The AUTOFIT tool of SLIM was used to provide the best non-linear least-squares LTE fit to the data using the Levenberg-Marquardt algorithm, which provides the value and uncertainty of the physical parameters for each molecular species \citep[see detailed description in][]{Martin2019}. The free parameters of the model are: molecular column density ($\textit{N}_{\rm tot}$), excitation temperature ($\textit{T}_{\rm ex}$), central radial velocity ($\upsilon_{\rm LSR}$), and full width half maximum (FWHM). Figure \ref{fig:CH3CONH2_E_spectra}$-$\ref{fig:H2NCONH2_spectra} present, in the order of increasing rest frequency, the unblended or slightly blended transitions of the molecules studied in this work. The emission of the other transitions as predicted by LTE are consistent with the fitted spectra, but they are severely blended with lines arising from other molecular species. The fitting parameters together with the derived column density and the molecular abundance with respect to molecular hydrogen, assuming $\textit{N}_{\rm H_2}$=1.35$\times$10$^{23}$ cm$^{-2}$ \citep{Martin2008}, of each molecule are summarised in Table \ref{tab:mol_parameters}. The transitions used for the LTE fits are listed in Table \ref{tab:mol_trans}.

    


\subsection{New detections}
\subsubsection*{Acetamide (CH\texorpdfstring{$_3$}{}C(O)NH\texorpdfstring{$_2$}{})}
Acetamide (CH$_3$C(O)NH$_2$) is an asymmetric top molecule that has an internal rotation of the methyl group, making its spectrum complicated. The dipole moments of CH$_3$C(O)NH$_2$ are $\mu_a$=1.14$\,$D and $\mu_b$=3.5$\,$D \citep{Hollis2006}. In this work, both a$-$ and b$-$type transitions at the ground torsional state were detected but b-type were predominant. For clarity, we adopt the same labelling convention J$^{\prime\prime}_{K^{\prime\prime}_a,K^{\prime\prime}_c} -$J$^{\prime}_{K^{\prime}_a,K^{\prime}_c}$ for the A$-$ and E$-$states of acetamide as \citet{Hollis2006}. Out of 79 unblended or slightly blended transitions, the energy range covered in the 34 E$-$state transitions (Figure \ref{fig:CH3CONH2_E_spectra}) and the 45 A$-$state transitions (Figure \ref{fig:CH3CONH2_A_spectra}) is E$\rm _u$=9.7$-$35.7$\,$K and E$\rm _u$=3.0$-$48.2$\,$K respectively. The best LTE fit finds good agreement in excitation conditions between A$-$ and E$-$state transitions. For E$-$state, $\textit{T}_{\rm ex}$=7.8$\pm$0.1$\,$K, $\upsilon_{\rm LSR}$=68.7$\pm$0.2$\,$km$\,$s$^{-1}$ and FWHM=21.6$\pm$0.4$\,$km$\,$s$^{-1}$ whilst for A$-$state, $\textit{T}_{\rm ex}$=7.4$\pm$0.2$\,$K, $\upsilon_{\rm LSR}$=68.7$\pm$0.3$\,$km$\,$s$^{-1}$, and FWHM=19.5$\pm$0.6$\,$km$\,$s$^{-1}$. The derived column density for E$-$ and A$-$state is $\textit{N}_{\rm tot}$=(8.1$\pm0.1)\times$10$^{13}$$\,$cm$^{-2}$ and $\textit{N}_{\rm tot}$=(3.4$\pm0.1)\times$10$^{13}$$\,$cm$^{-2}$ respectively, yielding an E/A ratio of 2.4$\pm$0.1 towards G+0.693 which is in agreement with the one, E/A = 1.9, derived from the colder more extended molecular gas towards Sgr B2(N) at an excitation temperature of 5.8$\,$K \citep{Remijan2022}. In contrast, an E/A ratio of 0.75 is obtained from the hot core region towards Sgr B2(N) at an excitation temperature of 170 K \citep{Belloche2013}. While the E/A ratio of CH$_3$C(O)NH$_2$ seems to vary with the excitation temperature, more detections of both E- and A-type CH$_3$C(O)NH$_2$ are needed to reach a conclusion. The total column density by summing up both states is $\textit{N}_{\rm tot}$=(1.15$\pm$0.02)$\times$10$^{14}$$\,$cm$^{-2}$, which translates to a molecular abundance of (8.5$\pm$0.1)$\times$10$^{-10}$.



\subsubsection*{Trans-N-methylformamide (CH\texorpdfstring{$_3$}{}NHCHO)}
N-methylformamide is an isomer of CH$_3$C(O)NH$_2$ and it exists in two forms, \textit{trans} and \textit{cis}, with the former being more stable \citep[by 466 cm$^{-1}$ or 666$\,$K; ][]{Kawashima2010}. In this work, both E$-$ and A$-$state transitions from the ground torsional state of trans$-$N$-$CH$_3$NHCHO are detected. Unlike CH$_3$C(O)NH$_2$, the same physical parameters as well as column densities were derived from both E$-$ and A$-$state trans$-$CH$_3$NHCHO. In total, 26 unblended or slightly blended transitions, 7 E$-$state transitions and 19 A$-$state transitions, were detected towards G+0.693 (Figure \ref{fig:CH3NHCHO_spectra}). The upper state energy ranges in E$\rm _u$=2.2$-$28.2 K. The best LTE fit gives $\textit{T}_{\rm ex}$=7.1$\pm$0.4$\,$K, $\upsilon_{\rm LSR}$=68.2$\pm$0.5$\,$km$\,$s$^{-1}$, FWHM=19$\pm$1$\,$km$\,$s$^{-1}$, and $\textit{N}_{\rm tot}$=(4.3$\pm0.4)\times$10$^{13}$$\,$cm$^{-2}$. This translates into a molecular abundance of (3.2$\pm$0.3)$\times$10$^{-10}$.

\subsection{Revisit}
\subsubsection*{Formamide (NH\texorpdfstring{$_2$}{}CHO)}
\label{formamide}
The detection of formamide (NH$_2$CHO) was reported towards G+0.693 by \citet{Zeng2018}. However, with the new dataset used in this study, it is confirmed that the NH$_2$CHO emission suffers from optical depth effect. Therefore, we have derived the column density of NH$_2$CHO using the optically thin $^{13}$C isotopologue. Six unblended or slightly blended transitions (E$\rm _u$ = 3.1$-$15.2$\,$K; Figure \ref{fig:NH213CHO_spectra}) are identified. The best-fit parameters obtained by AUTOFIT are $\textit{T}_{\rm ex}$=5.0$\pm$0.3$\,$K, $\upsilon_{\rm LSR}$=69.1$\pm$0.6$\,$km$\,$s$^{-1}$, FWHM=21$\pm$1$\,$km$\,$s$^{-1}$, and $\textit{N}_{\rm tot}$=(6.2$\pm$0.4)$\times$10$^{12}$$\,$cm$^{-2}$. Assuming $^{12}$C/$^{13}$C=40 measured in G+0.693 by Colzi et al. (in prep.), the inferred column density for NH$_2$CHO is (2.5$\pm$0.1)$\times$10$^{14}$$\,$cm$^{-2}$ and the corresponding molecular abundance for NH$_2$CHO is of (1.8$\pm$0.1)$\times$10$^{-9}$.

\subsection{Urea (NH\texorpdfstring{$_2$}{}C(O)NH\texorpdfstring{$_2$}{})}
Urea (NH$_2$C(O)NH$_2$), also known as carbamide, has a dipole moment of $\mu$=$\mu_b$=3.83$\,$D \citep{Brown1975}. A total of 15, among which 8 transitions have already been reported by \citet{Jimenez-Serra2020}, unblended or slightly blended transitions were detected towards G+0.693. The upper state energy ranges E$\rm _u$=2.9$-$25.7$\,$K. With $\upsilon_{\rm LSR}$ and FWHM fixed to 69$\,$km$\,$s$^{-1}$ and 20$\,$km$\,$s$^{-1}$, the best LTE fit gives $\textit{T}_{\rm ex}$=8.0$\pm$0.7$\,$K and $\textit{N}_{\rm tot}$=(7.1$\pm0.5)\times$10$^{12}$$\,$cm$^{-2}$. This translates into a molecular abundance of (5.2$\pm$0.5)$\times$10$^{-11}$, which is consistent to that reported by \citet{Jimenez-Serra2020}.

\section{Discussion}
\label{discussion}
\begin{figure*}
   \centering
  \includegraphics[width=\textwidth]{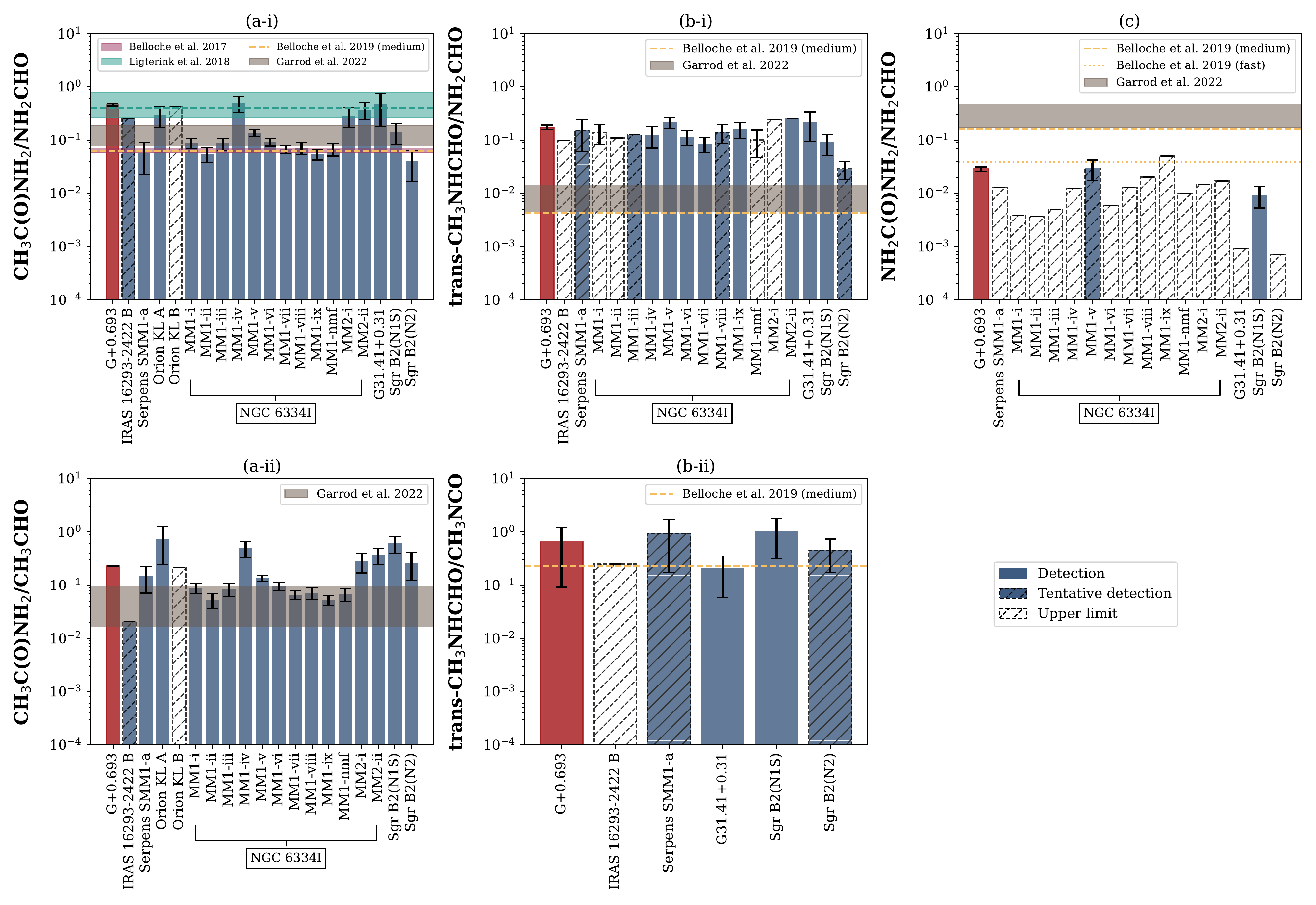}
   \caption{\textit{(a-i):} Ratio of CH$_3$C(O)NH$_2$/NH$_2$CHO. \textit{(a-ii):} Ratio of CH$_3$C(O)NH$_2$/CH$_3$CHO. \textit{(b-i):} Ratio of trans-N-CH$_3$NHCHO/NH$_2$CHO.  \textit{(b-ii):} Ratio of trans-N-CH$_3$NHCHO/CH$_3$NCO. \textit{(e):} Ratio of NH$_2$C(O)NH$_2$/NH$_2$CHO. The green, brown, and pink shaded area denote the results from laboratory experiment by \citet{Ligterink2018} and chemical models by \citet{Belloche2017} and \citet{Garrod2022}, respectively. The yellow dashed line and dotted line indicates the result from the best matching model, medium or fast, of \citet{Belloche2019} respectively. Observational data are taken from: G+0.693 (this work), IRAS 16293 B \citep{Coutens2016,Lykke2017,Ligterink2018}, Serpens SMM1 \citep{Ligterink2022}, Orion KL \citep{Cernicharo2016}, NGC 6334I \citep{Ligterink2020}, G31.41+0.31 \citep{Colzi2021}, Sgr B2(N1S) and (N2) \citep{Belloche2017,Belloche2019}.}
              \label{fig:mol_com}
    \end{figure*}

\begin{figure*}
   \centering
  \includegraphics[width=0.95\textwidth]{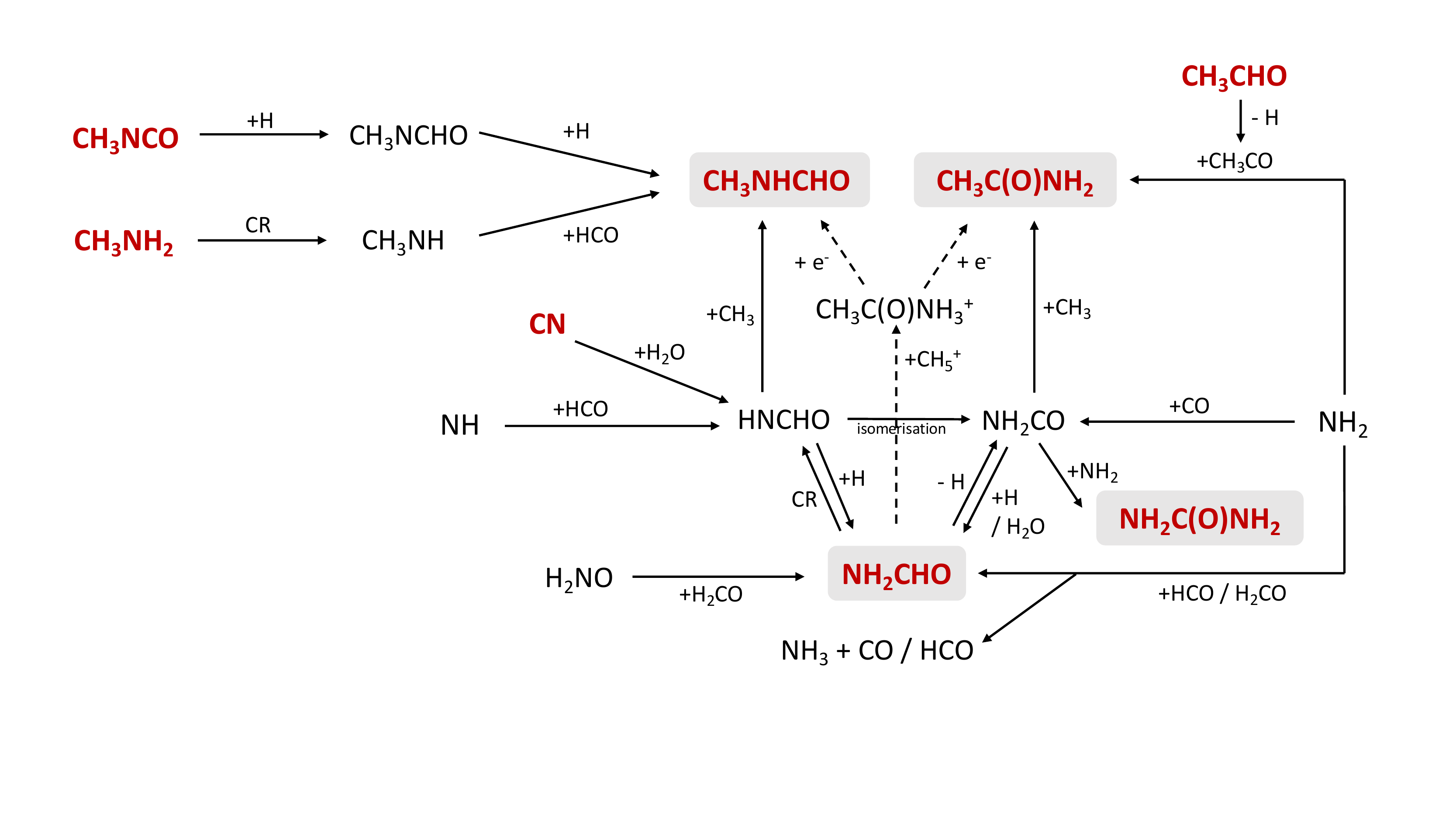}
       \vspace{-1.5cm}
   \caption{Summary of the proposed chemical formation routes of formamide, acetamide, trans-N-methylformamide, and urea in the ISM. Red denotes the molecular species detected towards G+0.693 molecular cloud and the grey shaded area denotes the amides species studied in this work. The solid and dashed arrows indicate surface chemistry reactions and gas-phase chemistry, respectively.}
              \label{fig:amides_network}%
    \end{figure*}

The amides detection presented in Section \ref{results} has not only expanded the number of species identified in G+0.693, but also offers an opportunity to study the chemical processes leading to amides formation in a region unaffected by star formation. Although G+0.693 is located in the same molecular cloud complex as the high-mass star-forming region Sgr B2(N), the kinematic and chemical study of \citet{Zeng2020} showed that the overall distribution of molecular gas towards G+0.693 depicted a different morphology and structure to that of Sgr B2(N). Furthermore, observational signatures of a cloud-cloud collision have been detected at small scales (0.2 pc), which agrees well with previous observations at larger scales in the Sgr B2 region as reported in literature \citep[e.g.][]{Hasegawa1994,Tsuboi2015}. As a consequence, shocks associated with the cloud-cloud collision are expected to sputter efficiently the icy grain mantles in G+0.693, subsequently ejecting materials formed on grains into the gas phase including complex organic molecules. Note that G+0.693 is located at a about $\sim$2.4 pc away in projected distance from the Sgr B2(N) high-mass star-forming region \citep[assuming a distance of 8.178 kpc for the Galactic Centre;][]{Abuter2019}, distant enough to make it unlikely that the star-formation activity of the Sgr B2(N) star cluster affects the chemistry of the G+0.693 cloud. Recent study of deuteration fractionation towards this source proposed that G+0.693 is on the verge of star formation \citep[][]{Colzi2022} i.e. under prestellar conditions, which is likely due to the event of a cloud-cloud collision. All this makes the conditions driving the chemistry of G+0.693 differ substantially from typical star-forming sources such as the nearby Sgr B2(N) hot core since the lack of protostellar heating implies that the thermal evaporation of ices is not yet playing a relevant role. Because amides have mostly been studied in star-forming regions, a comparison between these regions and G+0.693 could be useful to provide better constraints on the formation routes of amide species. In the following sections, comparison of molecular abundance of NH$_2$CHO with respect to H$_2$ is presented in Fig. \ref{fig:for_com} whilst molecular ratios with respect to NH$_2$CHO derived towards G+0.603 are are compared to those obtained from other interstellar sources (see Fig. \ref{fig:mol_com}). The comparison are made primarily towards star-forming sources in which the inventory of amides has been investigated in detail. 
The possible formation mechanisms of amide species, as summarised in Figure \ref{fig:amides_network}, are discussed along with available laboratory experiments, chemical models, and theoretical studies. To clarify, the term fast, medium, and slow models mentioned in the following section is referred to the model with warm-up timescales of 5 $\times$ 10$^4$ yr, 2 $\times$ 10$^5$ yr, and 1 $\times$ 10$^6$ yr, respectively as they are defined in \citet{Belloche2017,Belloche2019} and \citet{Garrod2022}.

\subsection{Formamide}

NH$_2$CHO is the smallest molecule to contain a peptide bond and hence often considered as a potential precursor of a wide variety of organic compounds essential to life. As reviewed by \citet{Lopez-Sepulcre2019}, both theoretical and experimental studies have investigated the formation mechanism of NH$_2$CHO. However, it is still strongly debated whether its formation is governed by gas-phase reactions or grain-surface chemistry. On grain surfaces, the popular route of successive HNCO hydrogenation (reaction 1 and 2) was ruled out by the experimental work of \citet{Noble2015} but disputed with the dual-cyclic mechanism of H-atom abstraction and addition reactions by \citet{Haupa2019}. 

\begin{align}
 \ce{
 HNCO + H &-> NH_2CO \\
 NH_2CO + H &-> NH_2CHO
 }
\end{align}

Another frequently studied formation route on the grain surface is the radical-radical recombination of NH$_2$ and HCO (reaction 3). It should proceed without any activation barrier; however, the two radicals are required to be in close proximity and a competing reaction (reaction 4) could lead back to the formation of NH$_3$ and CO \citep{Rimola2018,Chuang2022}. 

\begin{align}
 \ce{
 NH_2 + HCO &-> NH_2CHO \\
 NH_2 + HCO &-> NH_3 + CO
 }
\end{align}

Similar to reaction (3), a competing H-abstraction reaction is expected for radical-molecule reaction proposed by \citet{Fedoseev2016}:

\begin{align}
 \ce{
 NH_2 + H_2CO &-> NH_2CHO + H \\
 NH_2 + H_2CO &-> NH_3 + HCO
 }
\end{align}

Other surface reactions involving different precursors have also been proposed for the NH$_2$CHO formation. Through quantum chemical computations, \citet{Rimola2018} suggested a series of reactions forming NH$_2$CHO started by CN:

\begin{align}
 \ce{
 CN + H_2O &-> HNCOH \\
 HNCOH &-> NH_2CO \\
 NH_2CO + H_2O &-> NH_2CHO + OH
 }
\end{align}

Laboratory experiments performed by \citet{Dulieu2019} demonstrated that NH$_2$CHO formation is possible from barrierless hydrogenation of NO followed by the radical-radical addition reaction with H$_2$CO (reaction 10). However, under the conditions of typical molecular cloud i.e. at 10 K, NO and H$_2$CO are not mobile on the grain surface and hence their proximity represents the limiting factor of this formation route. 

\begin{align}
 \ce{
 H_2NO + H_2CO &-> NH_2CHO + OH
 }
\end{align}
In the gas phase, reaction (5) has been studied theoretically as a gas-phase pathway \citep{Barone2015,Skouteris2017}. However, its feasibility was challenged by \citet{Song2016} and more recently by \citet{Douglas2022}. The latter reported a significant energy barrier of the reaction and concluded that it is not an important source of NH$_2$CHO at low temperatures in interstellar environments. Note that at high temperatures, \citet{Quenard2018} also showed that the formation of NH$_2$CHO is dominated by grain surface reactions, thus its gas phase formation is not sufficient to reproduce the observations.

\begin{figure}
   \centering
  \includegraphics[width=0.48\textwidth]{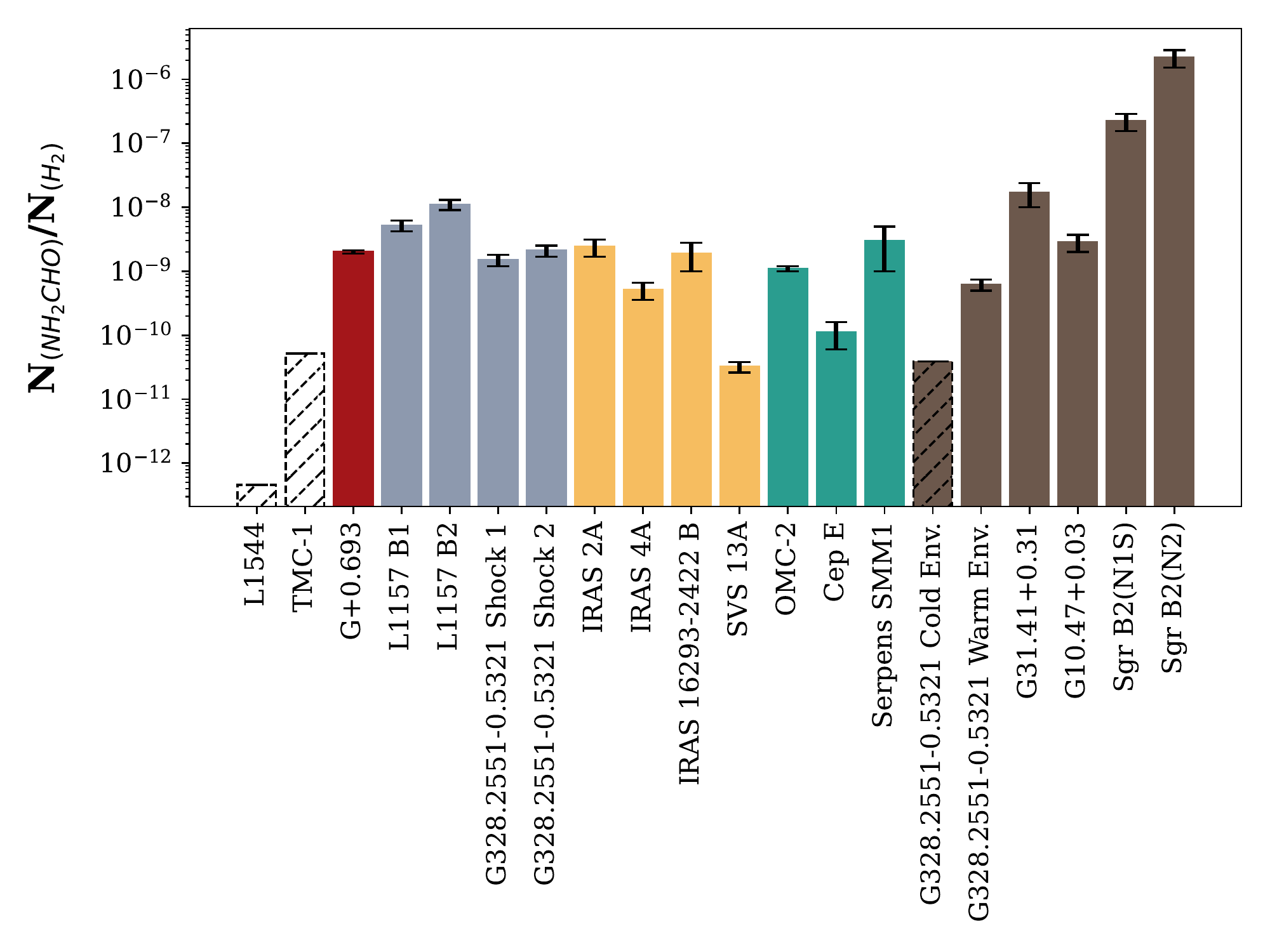}
      \vspace{-0.5cm}
   \caption{Bar plot of NH$_2$CHO abundance with respect to H$_2$ towards various interstellar regions. Sources with different physical conditions are represented in different colours: red = molecular cloud; grey = shocked regions; yellow = low-mass star-forming sources; green = intermediate star-forming sources; brown = high-mass star-forming sources. Upper limits of NH$_2$CHO are denoted in striped bar. Data are taken from: G+0.693 (this work), L1157 \citep{Mendoza2014}, IRAS 2A and IRAS 4A \citep{Taquet2015}, IRAS 16293 B \citep{Martin-Domenech2017}, L1544, TMC-1, SVS 13A, OMC-2, and Cep E \citep{Lopez-Sepulcre2015}, Serpens SMM1 \citep{Ligterink2022}, G31.41+0.31 \citep{Colzi2021}, G10.47+0.03 \citep{Gorai2020}, G328.2551-0.5321 \citep{Bouscasse2022}, Sgr B2 (N1S) and (N2)) \citep{Belloche2017,Belloche2019}. For Sgr B2 (N1S) and (N2), H$_2$ column density of 1.31$\times$10$^{25}$ cm$^{-2}$ and 1.6$\times$10$^{24}$ cm$^{-2}$ respectively, are adapted from \citet{Bonfand2017}.}
              \label{fig:for_com}%
    \end{figure}

In Figure \ref{fig:for_com}, the molecular abundance of NH$_2$CHO with respect to H$_2$ derived towards G+0.693 is compared to a diverse set of interstellar regions, including pre-stellar sources (L1544 and TMC-1), shocked regions (G+0.693, L1157, and G328.2551-0.5321 shock 1,2), low-mass (IRAS 2A, IRAS 4A, IRAS 16293-2422 B, SVS 13A), intermediate-mass (OMC-2, Cep E, Serpens SMM1), and high-mass star-forming sources (G31.41+0.31, G10.47+0.03, G328.2551-0.5321, Sgr B2N). To simplify, only sources with determined H$_2$ column densities were selected. In line with the conclusions of \citet{Colzi2021} and \citet{Ligterink2022}, the lack of detection towards pre-stellar sources and the comparable abundance of NH$_2$CHO between shocked regions and star-forming regions reveal that NH$_2$CHO is commonly formed in the ice mantles of grains and released back to gas-phase via non-thermal/thermal processes. In particular, the NH$_2$CHO abundance in G+0.693 is comparable, within a factor of 6, to the ones obtained in the shocked region L1157, G328.2551-0.5321 shock 1 and 2, suggesting that molecules detected towards G+0.693 have been freshly desorbed from the dust mantles as a result of grain-sputtering. The large abundance of NH$_2$CHO found in high-mass star-forming regions suggests that its formation could be boosted by the warm up of the grains due to star formation. However, this seems to be inconsistent with the constant abundance ratios, indicating that the formation might partly be occcuring during the pre-stellar phase as supported by recent chemical modelling \citep{Garrod2022}. In addition, relatively lower abundance of NH$_2$CHO in low-mass and intermediate-mass star-forming regions hints at a difference in the heating timescale.  

\subsection{Acetamide}
\label{ace}
The formation of CH$_3$C(O)NH$_2$ has been suggested to be dependent on the production of NH$_2$CHO \citep[e.g.][]{Hollis2006}. From an observational point of view, the column density of CH$_3$C(O)NH$_2$ is found to tightly correlate with that of NH$_2$CHO \citep[e.g.][]{Colzi2021,Ligterink2022}. The origin of this correlation could indicate a direct chemical link or the same response to the environmental conditions such as temperature for HNCO and NH$_2$CHO \citep{Quenard2018}. 

Chemically, the ion-molecule reaction (reaction 11 and 12)between protonated methane (CH$_5^+$) and neutral NH$_2$CHO followed by electron recombination is considered to be the most efficient process among various gas phase reactions proposed for CH$_3$C(O)NH$_2$ formation \citep{Halfen2011, Redondo2014}. 
\begin{align}
 \ce{
 CH_5^+ + NH_2CHO  &-> CH_3C(O)NH_3^+ + H_2 \\
 CH_3C(O)NH_3^+ + e^- &->CH_3C(O)NH_2 + H
 }
\end{align}
On the surface of dust grains, CH$_3$C(O)NH$_2$ is considered to be predominately formed by the addition of CH$_3$ to NH$_2$CO radical (reaction 13) and the latter links CH$_3$C(O)NH$_2$ and NH$_2$CHO through several possible reactions.
\begin{align}
 \ce{
 CH_3 + NH_2CO &-> CH_3C(O)NH_2
 }
\end{align}
 
In recent chemical models, this radical is considered to be formed by H-abstraction from NH$_2$CHO, addition of H to HNCO, and/or radical addition of NH$_2$ + CO \citep[][]{Belloche2017,Belloche2019,Garrod2022}. In the case of H-abstraction from NH$_2$CHO, \citet{Belloche2017} found a CH$_3$C(O)NH$_2$/NH$_2$CHO ratio of 0.06$-$0.07 (denoted as pink shaded area in Figure \ref{fig:mol_com}-(a-i)). In contrast, \citet{Garrod2022} determined the ratio of 0.08$-$0.19 (denoted as brown shaded area in Figure \ref{fig:mol_com}-(a-i)) by considering both the H addition to HNCO and reaction of NH$_2$ with CO in the bulk ice for the NH$_2$CO radical production. The derived CH$_3$C(O)NH$_2$/NH$_2$CHO ratio in G+0.693 is 0.43$\pm$0.03 which is at least a factor of two higher than the model predicted ratios. We note that these models are for hot cores where thermal desorption dominates the ejection of molecules to gas phase from grains. But since the formation of CH$_3$C(O)NH$_2$ is likely dominated by grain surface chemistry, some insights might be provided by making the comparison between the observational results in G+0.693 and these chemical models. In fact, several high-mass star-forming regions such as Orion KL, NGC 6334I (MM1-iv, MM2-i, and MM2-ii), and G31.41+0.31 also show CH$_3$C(O)NH$_2$/NH$_2$CHO ratios close to that found in G+0.693 (see Figure \ref{fig:mol_com}-(a-i)). Additionally, this ratio determined towards G+0.693 is consistent with the results from laboratory experiment in which CH$_4$–HNCO ice mixtures at 20 K were irradiated with far-UV photons to produce the radicals required for the formation of the amides \citep[green shaded area in \ref{fig:mol_com}-(a-i);][]{Ligterink2018}. The resulting CH$_3$C(O)NH$_2$/NH$_2$CHO ratio is 0.4${^{+0.39}_{-0.14}}$ and the NH$_2$CO radical is proposed to form via the radical addition of NH$_2$ + CO. The good match suggests that the formation pathway of both NH$_2$CHO and CH$_3$C(O)NH$_2$ on dust grains in G+0.693 are similar to the chemical processes mimicked on laboratory ices i.e. NH$_2$CHO formed via reaction (3) and CH$_3$C(O)NH$_2$ formed via reaction (13), of which the NH$_2$CO radical is the product of NH$_2$ + CO. Therefore, both species might not be directly connected in chemistry but likely share a common chemical origin, similar to the scenario explained by \citet{Garrod2022} for HNCO and NH$_2$CO. Since G+0.693 is shielded from direct UV irradiation, the key intermediate radicals are likely produced by cosmic ray-induced UV field instead. It is noteworthy that the reaction of NH$_2$ + HCO is found to be negligible compared with both the abstraction of hydrogen from NH$_2$CHO and the addition of H to HNCO in chemical models by setting a typical value of activation energy barrier to 2500 K \citep{Belloche2019}. Therefore, constraining the activation barriers and rate coefficient of the processes that lead to NH$_2$CO, the key intermediate, is one essential step to better elucidate the formation of CH$_3$C(O)NH$_2$.

In the literature, there are other alternative mechanisms proposed for the formation of  CH$_3$C(O)NH$_2$. For instance, one involves addition of NH$_2$ radical to CH$_3$CO (reaction (14)), which may also be important to form CH$_3$C(O)NH$_2$ on dust grain surfaces \citep[][]{Belloche2017,Ligterink2018,Belloche2019,Garrod2022}. 
\begin{align}
 \ce{
 NH_2 + CH_3CO &-> CH_3C(O)NH_2
 }
\end{align}
The CH$_3$CO could be formed by H-abstraction of acetaldehyde (CH$_3$CHO) which is similar to the H abstraction of NH$_2$CHO. Towards G+0.693, the abundance of CH$_3$CHO is derived to be 3.7$\times$10$^{-9}$ \citep{Sanz-Novo2022} and it is approximately as abundant as NH$_2$CHO. Therefore the formation pathway of NH$_2$ + CH$_3$CO seems to be feasible in G+0.693. In the study of \citet{Ligterink2018}, CH$_3$CHO was not detected in the laboratory experiments which raised the speculation of this formation route to proceed on ices, but it could not be completely ruled out. In recent chemical models \citep[see brown shaded area in Figure \ref{fig:mol_com}-(a-ii)][]{Belloche2019,Garrod2022}, this formation route is included as modest contribution to form CH$_3$C(O)NH$_2$ and it is found to become important at around T=50$\,$K in a slow warm-up model. However, the ratio of CH$_3$C(O)NH$_2$/CH$_3$CHO does not vary significantly (within a factor of 5) between G+0.693, whose dust temperature typically lie below 20 K \citep[][]{Rodriguez-Fernandez2004,Guzman2015}, and star-forming sources (Figure \ref{fig:mol_com}-(a-ii)). The comparison here points towards the idea that reaction (14) may have contribution to the formation of CH$_3$C(O)NH$_2$ but not the dominant mechanism on the grain surface. 

\subsection{Trans-N-methylformamide}
As for the trans-N-CH$_3$NHCHO/NH$_2$CHO ratio, a scatter of less than a factor of 3, except Sgr B2(N2), is found across various sources. The consistent results between G+0.693 and star-forming sources indicates that the tentative correlation between trans-N-CH$_3$NHCHO and NH$_2$CHO hold also at an early stage of star formation and the formation route of trans-N-CH$_3$NHCHO could be similar among all these sources. 

On grain surfaces, trans-N-CH$_3$NHCHO is considered to be linked to NH$_2$CHO by the reaction of HNCHO + CH$_3$ \citep[][]{Belloche2017,Ligterink2018,Garrod2022}: 
\begin{align}
 \ce{
HNCHO + CH_3 &-> CH_3NHCHO }
\end{align}
The HNCHO radical is an isomer of NH$_2$CO. Although it is unclear how these two species differ in term of reactivity, HNCHO should be more stable than NH$_2$CO since the molecular orbital of the unpaired electron held by the nitrogen atom could be stabilised by hyperconjugation by the C$-$O $\pi$ bond (J. Garcia de la Concepcion, private communication). HNCHO can be formed via the radical addition of NH + CHO, but H-addition yielding NH$_2$CHO is expected to occur subsequently. Whilst H-abstraction from NH$_2$CHO seems to strongly favour the production of NH$_2$CO \citep{Belloche2017}, cosmic ray-induced photo-dissociation of NH$_2$CHO as another possible formation route of HNCHO \citep[][]{Belloche2017,Ligterink2018}. Consequently, the production of trans-N-CH$_3$NHCHO largely depend on the availability of HNCHO which may not be formed efficiently or could be easily consumed in the formation of other molecules. As indicated in brown shaded area and yellow dashed line in the upper-middle panel of Figure \ref{fig:mol_com}-(b-i), chemical models yield a ratio of trans-N-CH$_3$NHCHO/NH$_2$CHO nearly an order of magnitude lower than the observational values.

In the chemical model of \citet{Belloche2017}, trans-N-CH$_3$NHCHO is proposed to also form via hydrogenation of methyl isocyanate (CH$_3$NCO):
\begin{align}
 \ce{
CH_3NCO + H &-> CH_3NHCO \\
CH_3NHCO + H &-> CH_3NHCHO 
}
\end{align}
However, \citet{Ligterink2018} have cast doubt on this pathway based on the non-detection of trans-N-CH$_3$NHCHO in laboratory ice experiments whilst CH$_3$NCO was formed abundantly. With the secure detection of trans-N-CH$_3$NHCHO for the first time in ISM towards Sgr B2 (N1S), chemical models of \citet{Belloche2019} indicated that hydrogenation dominates in trans-N-CH$_3$NHCHO formation. The best matching model resulted in a trans-N-CH$_3$NHCHO/CH$_3$NCO ratio of 0.23 which is denoted as yellow dashed line in Figure \ref{fig:mol_com}-(b-ii). The ratio obtained from observations towards different sources is consistent within the uncertainties of the modelling results, implying that the hydrogenation of CH$_3$NCO is one plausible source of trans-N-CH$_3$NHCHO production on grain surfaces. The same chemical network of trans-N-CH$_3$NHCHO has also been adopted in \citet{Garrod2022}, but the trans-N-CH$_3$NHCHO/CH$_3$NCO ratio ranged between 7 to 24. The authors noted that the activation-energy barriers of the grain-surface reaction for CH$_3$NCO production are poorly defined, which results in substantially under-production of CH$_3$NCO.

Alternatively, \citet{Frigge2018} demonstrated the formation of trans-N-CH$_3$NHCHO via reaction (18) in a mixture of methylamine (CH$_3$NH$_2$) and CO ices where the HNCH$_3$ radical is created from CH$_3$NH$_2$ by irradiation with energetic electrons i.e. by simulating cosmic rays. 
\begin{align}
 \ce{
HNCH_3 + HCO &-> CH_3NHCHO 
}
\end{align}
Using the column density of CH$_3$NH$_2$ derived towards G+0.693 by \citet{Zeng2018}, the CH$_3$C(O)NH$_2$/CH$_3$NH$_2$ ratio is $\sim$0.01. The comparison of this ratio is currently not available as detections of CH$_3$NH$_2$ are very limited in the literature, especially towards the same emitting regions with similar spatial resolution. For instance, CH$_3$NH$_2$ has been reported towards Sgr B2(N) using single-dish observations \citep{Belloche2013} whilst the emission of trans-N-CH$_3$NHCHO is detected towards Sgr B2(N1S) and Sgr B2(N2) with ALMA \citep{Belloche2019}. Furthermore, high abundance of trans-N-CH$_3$NHCHO has been reported in G31.41 but the existence of CH$_3$NH$_2$ is not yet clear \citep{Ohishi2019}. Increasing the number of trans-N-CH$_3$NHCHO detections together with NH$_2$CHO, CH$_3$NCO, and CH$_3$NH$_2$ will better constrain the chemical models and unravel the contributions of different formation pathways.

\subsection{Urea}
Up to date, G+0.693 is the only other source besides Sgr B2(N1S) where NH$_2$C(O)NH$_2$ is securely detected \citep{Belloche2019,Jimenez-Serra2020}. In this work, the derived NH$_2$C(O)NH$_2$/NH$_2$CHO ratio is 0.030$\pm$0.003 which is consistent to the tentative detection towards NGC 6334I MM1-v \citep[$\sim$0.03][]{Ligterink2020} but a factor of $\sim$3 higher than the one obtained in Sgr B2(N1S) \citep[0.009;][]{Belloche2019}. The comparison is shown in Figure \ref{fig:mol_com}-(c). 

The chemical model by \citet{Belloche2019} considered NH$_2$C(O)NH$_2$ is formed directly by reaction (19) in the grain-surface ices:
\begin{align}
 \ce{
NH_2CO + NH_2 &-> NH_2C(O)NH_2 }
\end{align}
It has also been suggested to be one of the most promising formation pathways for NH$_2$C(O)NH$_2$ in both experimental and computational studies \citep{Ligterink2018,Slate2020}. However, as mentioned in Section \ref{ace}, the formation of NH$_2$CO radical is still a subject of debate. Unlike the case of CH$_3$C(O)NH$_2$, there is no information about the relative abundance between NH$_2$C(O)NH$_2$ and NH$_2$CHO in the laboratory experiment by \citet{Ligterink2018}. Thus, it is unclear if a consistent result can be found between observation towards G+0.693 and the experimental studies for NH$_2$C(O)NH$_2$. On the other hand, the NH$_2$C(O)NH$_2$/NH$_2$CHO ratio derived in G+0.693 is marginally consistent with the value estimated by the fast model of \citet{Belloche2019} ($\sim$0.039; see yellow dotted line in Figure \ref{fig:mol_com}-(c)). However, the ratio from the medium model of \citet{Garrod2022}, which appears to reproduce best other amide species, is at least a factor of a few higher than observed. This indicates that the formation pathways considered may be too efficient on the grains. It was pointed out by the authors that a higher barrier to the production of NH$_2$ from H-abstraction of ammonia and an efficient destruction route in the gas phase would help to ameliorate the disagreement. Furthermore, the fast model of \citet{Belloche2019} revealed that the solid-phase urea is seen to be produced most strongly after around 55 K, which is the temperature of rapid production and diffusion of NH$_2$ radical. Owing to the fact that the dust temperature in G+0.693 is low, a higher ratio of NH$_2$C(O)NH$_2$/NH$_2$CHO derived towards G+0.693 in comparison to Sgr B2 (N1S) suggests that either another formation route might at play at lower temperatures, or a destruction route at high temperature in the gas phase. 

Besides radicals, formation pathways involving charged species have been put forward to produce NH$_2$C(O)NH$_2$ in the ISM. Theoretical calculation by \citet{Jeanvoine2019} proposed that the reaction of protonated hydroxylamine (NH$_2$OH$_2^+$) with neutral NH$_2$CHO could be one alternative route. Whilst the authors concluded that this pathway would not be readily accessible in the gas phase without certain conditions being met, the presence of ice mantle may facilitate the overall reaction process. The neutral hydroxylamine (NH$_2$OH) is detected with an abundance of 2.1$\times$10$^{-10}$ \citep{Rivilla2020} towards G+0.693, but it is unclear how efficient its cation can be formed on grain surfaces. In addition, the ion-molecule reaction between N-protonated isocyanic acid (H$_2$NCO$^+$) and NH$_3$ have considered to be responsible for the NH$_2$C(O)NH$_2$ formation, both in the gas phase and solid state \citep[][and references therein]{Slate2020}. With the H$_2$NCO$^+$ abundance of $\sim$10$^{-11}$ derived towards G+0.693 \citep{Rodriguez-Almeida2021b}, which is lower than the one derived for NH$_2$C(O)NH$_2$, this route therefore does not appears to be viable in G+0.693. 

At present, the formation of NH$_2$C(O)NH$_2$ in the ISM is best described by the particular radical-radical reaction of NH$_2$CO and NH$_2$. Further exploration of not only ion-molecule reactions on the grain surfaces but also gas-phase formation route of NH$_2$C(O)NH$_2$, which has not been given a great deal of attention, are required to improve our understanding of the NH$_2$C(O)NH$_2$ chemistry in the ISM.

\section{Conclusions}
\label{conclusions}
In this work the amide inventory of the Galactic Centre molecular cloud G+0.693 is studied. As a molecular cloud that is proposed to be on the verge of star formation, G+0.693 is a well-suited testbed to study the interstellar chemistry at an early evolutionary stage. With high-sensitivity observations, the robust detection of all NH$_2$CHO, CH$_3$C(O)NH$_2$, trans-N-CH$_3$NHCHO, and NH$_2$C(O)NH$_2$ is reported, for the first time, towards a region that does not show any star-formation activities. Abundances of amides, derived with respect to H$_2$ and NH$_2$CHO, are compared with those obtained across various astronomical sources. The column density ratios of amide species with respect to NH$_2$CHO do not vary more than an order of magnitude regardless of physical conditions of different sources. This suggests that they are either chemically linked to NH$_2$CHO or share a common chemical origin with NH$_2$CHO. Based on the current available literature, the formation pathway of the amide species discussed in the study are generally dominated by grain surface chemistry but the specific chemical process(es) are unclear. However the rather constant abundance ratios found between G+0.693 and star-forming regions suggests that the amide related chemistry, which drives the observed relative abundances, are already set in early evolutionary stages of molecular cloud, and star formation process neither destroys these ratios, nor enhances them through the warm up process. One way to gain insight would be by constraining the formation route of the key precursor, the NH$_2$CO radical. Considering that several precursors have been detected in G+0.693, attention should not be limited to NH$_2$CHO but should be to alternative routes involving other molecular species. The prolific amide inventory in G+0.693 confirms it is one of the largest molecular repositories of very complex molecules in the Galaxy.

%

\section*{Acknowledgements}


We would like to thank Miguel Sanz-Novo and Juan Garc\'ia de la Concepci\'on for helpful discussions. We thank the anonymous referee for a constructive report that improved the quality of this manuscript. We are grateful to the IRAM 30m telescope staff for their help during the different observing runs. IRAM is supported by INSU/CNRS (France), MPG (Germany) and IGN (Spain). We thank Arnaud Belloche and Vadim Ilyushin for providing us the spectroscopic entry of CH$_3$C(O)NH$_2$ and trans-N-CH$_3$NHCHO. V.M.R. has received support from the project RYC2020-029387-I funded by MCIN/AEI /10.13039/501100011033, and from the Comunidad de Madrid through the Atracci\'on de Talento Investigador Modalidad 1 (Doctores con experiencia) Grant (COOL: Cosmic Origins Of Life; 2019-T1/TIC-5379). I.J.-S., J.M.-P. and L. C. acknowledge support from grant No. PID2019-105552RB-C41 by the Spanish Ministry of Science and Innovation/State Agency of Research MCIN/AEI/10.13039/501100011033 and by "ERDF A way of making Europe". P.d.V. and B.T. thank the support from the Spanish Ministerio de Ciencia e Innovacion (MICIU) through project PID2019-107115GB-C21. B.T. also acknowledges the Spanish MICIU for funding support from grant PID2019-106235GB-I00.

\section*{Data Availability}
The data underlying this article will be shared on reasonable request to the corresponding author.



\bibliographystyle{mnras}
\bibliography{ref} 




\appendix

\section{Molecular spectroscopy}
   \begin{table*}
   \centering
    \caption{Line list and spectroscopic references for the molecules studied in this work}
        \label{tab:mol_spec}
    \begin{threeparttable}
    \begin{tabular}{ccccc}
         \hline
         \hline
         Molecule & Catalogue & Entry & Date & Reference \\
         \hline
         NH$_2$CHO & CDMS & 45512 & April 2013 & \makecell{(1), (2), (3), (4),\\(5), (6), (7), (8)} \\
         NH$_2^{13}$CHO & CDMS & 46512 & April 2013 & (3), (6), (7), (8), (9) \\
         CH$_3$C(O)NH$_2$ & This work & $-$ & December 2020 & (10), (11)\\
         trans-N-CH$_3$NHCHO & This work & $-$ & December 2020 & (11)\\
         NH$_2$C(O)NH$_2$ & CDMS & 60517 & October 2017 & (12), (13), (14), (15)\\
         \hline
    \end{tabular}
    \begin{tablenotes}[flushleft]
        \small
            \item \textbf{Notes.} The species labelled with MADCUBA were imported into \textsc{madcuba}, using the spectroscopic works indicated in the table.
            \item \textbf{References.} (1)\citet{Kukolich1971}; (2)\citet{Hirota1974}; (3)\citet{Gardner1980}; (4)\citet{Moskienko1991}; (5)\citet{Vorobeva1994}; (6)\citet{Blanco2006}; (7)\citet{Kryvda2009}; (8)\citet{Motiyenko2012}; (9)\citet{Stubgaard1978}; (10)\citet{Ilyushin2004}; (11)\citet{Belloche2017}; (12)\citet{Remijan2014}; (13)\citet{Brown1975}; (14)\citet{Kasten1986}; (15)\citet{Kretschmer1996}
        \end{tablenotes}
     \end{threeparttable}
   \end{table*}
%

    \clearpage
    \onecolumn

\begin{longtable}[c]{cccccc}
    \caption{List of detected unblended or slightly blended transitions of the amides analysed in this work. The columns indicate the frequency, quantum numbers, logarithm of the integrated intensity at 300 K, energy of the upper levels of each transition (E$\rm_{u}$), and information about the possible blending by other identified or unidentified (U) species towards G+0.693. Asterisk ($^*$) indicates the transitions detected in previous studies towards G+0.693. \label{mol_spec}} \\
        \hline
        \hline
        \textbf{Molecule} & \textbf{Frequency} & \textbf{Transition} & \textbf{logI(300K)} & \textbf{E$\rm_{u}$} & \textbf{Blending} \\
            & (GHz) & (J$^{\prime\prime}_{K^{\prime\prime}_a,K^{\prime\prime}_c} -$J$^{\prime}_{K^{\prime}_a,K^{\prime}_c}$ E/A)  & (nm$^{2}$MHz) & (K) & \\
        \hline
        \endfirsthead
        \multicolumn{4}{c}%
        {\tablename\ \thetable\ -- \textit{Continued from previous page}} \\
        \hline
        \textbf{Molecule} & \textbf{Frequency} & \textbf{Transition} & \textbf{logI(300K)} & \textbf{E$\rm_{u}$} & \textbf{Blending} \\
           & (GHz) & (J$^{\prime\prime}_{K^{\prime\prime}_a,K^{\prime\prime}_c} -$J$^{\prime}_{K^{\prime}_a,K^{\prime}_c}$ E/A)  & (nm$^{2}$MHz) & (K) & \\
        \hline
        \endhead
        \hline \multicolumn{4}{r}{\textit{Continued on next page}} \\
        \endfoot
        \hline
        \endlastfoot
         CH$_3$C(O)NH$_2$ & 32.640032 & 3$_{0,3}-$2$_{1,2}$ E & -6.344 & 9.71 & unblended or blended with U\\
         CH$_3$C(O)NH$_2$ & 35.226603 & 4$_{3,2}-$3$_{3,1}$ E & -6.687 & 15.04 & unblended\\
         CH$_3$C(O)NH$_2$ & 39.338502 & 3$_{1,3}-$2$_{0,2}$ E & -6.137 & 9.74 & unblended or blended with U\\
         CH$_3$C(O)NH$_2$ & 40.755758 & 5$_{1,4}-$5$_{0,5}$ E & -6.285 & 16.67 & unblended\\
         CH$_3$C(O)NH$_2$ & 46.534292 & 4$_{0,4}-$3$_{1,3}$ E & -5.668 & 11.98 & unblended\\
         CH$_3$C(O)NH$_2$ & 77.317157 & 3$_{0,3}-$2$_{1,2}$ E & -6.476 & 13.45 & unblended\\
         CH$_3$C(O)NH$_2$ & 77.329953 & 7$_{0,7}-$6$_{1,6}$ E & -4.950 & 21.64 & unblended\\
         CH$_3$C(O)NH$_2$ & 77.330094 & 7$_{1,7}-$6$_{1,6}$ E & -6.473 & 21.64 & unblended\\
         CH$_3$C(O)NH$_2$ & 77.331133 & 7$_{0,7}-$6$_{0,6}$ E & -6.473 & 21.64 & unblended\\
         CH$_3$C(O)NH$_2$ & 77.331275 & 7$_{1,7}-$6$_{0,6}$ E & -4.950 & 21.64 & unblended\\
         CH$_3$C(O)NH$_2$ & 77.900349 & 3$_{3,0}-$2$_{2,0}$ E & -5.404 & 14.90 & unblended\\
         CH$_3$C(O)NH$_2$ & 87.604693 & 8$_{0,8}-$7$_{1,7}$ E & -4.785 & 25.85 & shoulder of HNCO and HN$^{13}$CO\\
         CH$_3$C(O)NH$_2$ & 87.604710 & 8$_{1,8}-$7$_{1,7}$ E & -6.252 & 25.85 & shoulder of HNCO and HN$^{13}$CO\\
         CH$_3$C(O)NH$_2$ & 87.604835 & 8$_{0,8}-$7$_{0,7}$ E & -6.252 & 25.85 & shoulder of HNCO and HN$^{13}$CO\\
         CH$_3$C(O)NH$_2$ & 87.604852 & 8$_{1,8}-$7$_{0,7}$ E & -4.785 & 25.85 & shoulder of HNCO and HN$^{13}$CO\\
         CH$_3$C(O)NH$_2$ & 93.680900 & 6$_{3,4}-$5$_{2,3}$ E & -5.187 & 22.36 & unblended\\
         CH$_3$C(O)NH$_2$ & 97.810915 & 8$_{1,7}-$7$_{2,6}$ E & -4.771 & 29.32 & blended with Ga-n-C$_3$H$_7$OH\\
         CH$_3$C(O)NH$_2$ & 97.812712 & 8$_{2,7}-$7$_{2,6}$ E & -6.242 & 29.32 & blended with Ga-n-C$_3$H$_7$OH\\
         CH$_3$C(O)NH$_2$ & 97.823791 & 8$_{1,7}-$7$_{1,6}$ E & -6.243 & 29.32 & blended with H$_2$CCCHCN\\
         CH$_3$C(O)NH$_2$ & 97.825588 & 8$_{2,7}-$7$_{1,6}$ E & -4.771 & 29.32 & blended with H$_2$CCCHCN\\
         CH$_3$C(O)NH$_2$ & 97.893420 & $9_{0,9}-$8$_{1,8}$ E & -4.640 & 30.54 & blended with aGg$^{\prime}-$(CH$_2$OH)$_2$\\
         CH$_3$C(O)NH$_2$ & 97.893422 & $9_{1,9}-$8$_{1,8}$ E & -6.069 & 30.54 & blended with aGg$^{\prime}-$(CH$_2$OH)$_2$\\
         CH$_3$C(O)NH$_2$ & 97.893437 & $9_{0,9}-$8$_{0,8}$ E & -6.069 & 30.54 & blended with aGg$^{\prime}-$(CH$_2$OH)$_2$\\
         CH$_3$C(O)NH$_2$ & 97.893439 & $9_{1,9}-$8$_{0,8}$ E & -4.640 & 30.54 & blended with aGg$^{\prime}-$(CH$_2$OH)$_2$\\
         CH$_3$C(O)NH$_2$ & 98.113341 & $4_{4,0}-$3$_{3,0}$ E & -5.055 & 19.61 & unblended\\
         CH$_3$C(O)NH$_2$ & 107.988285 & $9_{1,8}-$8$_{2,7}$ E & -4.626 & 34.51 & unblended\\
         CH$_3$C(O)NH$_2$ & 107.990327 & $9_{2,8}-$8$_{1,7}$ E & -4.626 & 34.51 & unblended\\
         CH$_3$C(O)NH$_2$ & 108.190197 & $10_{0,10}-$9$_{1,9}$ E & -4.511 & 35.74 & unblended\\
         CH$_3$C(O)NH$_2$ & 108.190197 & $10_{,110}-$9$_{1,9}$ E & -5.912 & 35.74 & unblended\\
         CH$_3$C(O)NH$_2$ & 108.190199 & $10_{0,10}-$9$_{0,9}$ E & -5.912 & 35.74 & unblended\\
         CH$_3$C(O)NH$_2$ & 108.190200 & $10_{1,10}-$9$_{0,9}$ E & -4.511 & 35.74 & unblended\\
         CH$_3$C(O)NH$_2$ & 108.606162 & $8_{3,6}-$7$_{2,5}$ E & -4.784 & 32.27 & unblended\\
         CH$_3$C(O)NH$_2$ & 109.349123 & $5_{4,1}-$4$_{3,1}$ E & -5.266 & 21.54 & blended with aGg$^{\prime}-$(CH$_2$OH)$_2$\\
         CH$_3$C(O)NH$_2$ & 113.867594 & $5_{5,1}-$4$_{4,1}$ E & -4.993 & 23.30 & share with  CH$_3$C(O)NH$_2$ $7_{4,4}-$6$_{3,3}$ A\\
         CH$_3$C(O)NH$_2$ & 34.987857 & 6$_{5,2}-$6$_{4,3}$ A & -5.997 & 20.20 & blended with anti-C$_2$H$_5$NH$_2$\\
         CH$_3$C(O)NH$_2$ & 34.989515 & 4$_{1,3}-$4$_{0,4}$ A & -6.372 & 7.61 & blended with anti-C$_2$H$_5$NH$_2$\\
         CH$_3$C(O)NH$_2$ & 40.302903 & 3$_{1,2}-$2$_{2,1}$ A & -6.417 & 4.84 & unblended or blended with U\\
         CH$_3$C(O)NH$_2$ & 46.312309 & 4$_{0,4}-$3$_{1,3}$ A & -5.662 & 5.93 & unblended\\
         CH$_3$C(O)NH$_2$ & 46.449611 & 2$_{2,0}-$1$_{1,1}$ A & -6.262 & 3.03 & blended with H$_2$CCCHCN\\
         CH$_3$C(O)NH$_2$ & 46.450396 & 4$_{0,4}-$3$_{0,3}$ A & -6.844 & 5.93 & blended with H$_2$CCCHCN\\
         CH$_3$C(O)NH$_2$ & 46.473480 & 4$_{1,4}-$3$_{0,3}$ A & -5.658 & 5.93 & unblended\\   
         CH$_3$C(O)NH$_2$ & 74.172527 & 4$_{3,2}-$3$_{2,1}$ A & -5.479 & 9.00 & unblended or blended with U\\
         CH$_3$C(O)NH$_2$ & 77.199071 & 6$_{1,5}-$5$_{2,4}$ A & -5.136 & 14.56 & blended with U\\
         CH$_3$C(O)NH$_2$ & 77.320853 & 7$_{0,7}-$6$_{1,6}$ A & -4.951 & 15.58 & unblended\\
         CH$_3$C(O)NH$_2$ & 77.320920 & 7$_{1,7}-$6$_{1,6}$ A & -6.153 & 15.58 & unblended\\
         CH$_3$C(O)NH$_2$ & 77.321347 & 7$_{0,7}-$6$_{0,6}$ A & -6.153 & 15.58 & unblended\\
         CH$_3$C(O)NH$_2$ & 77.321414 & 7$_{1,7}-$6$_{0,6}$ A & -4.950 & 15.58 & unblended\\
         CH$_3$C(O)NH$_2$ & 77.435420 & 6$_{2,5}-$5$_{1,4}$ A & -5.133 & 14.56 & unblended or blended with U\\
         CH$_3$C(O)NH$_2$ & 82.338160 & 5$_{3,3}-$4$_{2,2}$ A & -5.341 & 12.59 & unblended\\
         CH$_3$C(O)NH$_2$ & 85.746007 & 6$_{2,4}-$5$_{3,3}$ A & -5.221 & 16.70 & blended with C$_2$H$_5$OH\\
         CH$_3$C(O)NH$_2$ & 87.586466 & 7$_{1,6}-$6$_{2,5}$ A & -4.938 & 18.76 & blended with H$_2$CCCHCN\\
         CH$_3$C(O)NH$_2$ & 87.623526 & 7$_{1,6}-$6$_{1,5}$ A & -6.098 & 18.76 & unblended\\
         CH$_3$C(O)NH$_2$ & 87.629757 & 7$_{2,6}-$6$_{1,5}$ A & -4.937 & 18.76 & unblended\\
         CH$_3$C(O)NH$_2$ & 87.632434 & 8$_{0,8}-$7$_{1,7}$ A & -4.784 & 19.79 & unblended\\
         CH$_3$C(O)NH$_2$ & 87.632443 & 8$_{1,8}-$7$_{1,7}$ A & -5.989 & 19.79 & unblended\\
         CH$_3$C(O)NH$_2$ & 87.632501 & 8$_{0,8}-$7$_{0,7}$ A & -5.989 & 19.79 & unblended\\
         CH$_3$C(O)NH$_2$ & 87.632510 & 8$_{1,8}-$7$_{0,7}$ A & -4.784 & 19.79 & unblended\\
         CH$_3$C(O)NH$_2$ & 97.905697 & $8_{1,7}-$7$_{2,6}$ A & -4.77 & 23.46 & unblended\\
         CH$_3$C(O)NH$_2$ & 97.906677 & $8_{2,7}-$7$_{2,6}$ A & -5.941 & 23.46 & unblended\\
         CH$_3$C(O)NH$_2$ & 97.911929 & $8_{1,7}-$7$_{1,6}$ A & -5.941 & 23.46 & unblended\\
         CH$_3$C(O)NH$_2$ & 97.912908 & $8_{2,7}-$7$_{1,6}$ A & -4.77 & 23.46 & unblended\\
         CH$_3$C(O)NH$_2$ & 97.943873 & $9_{0,9}-$8$_{1,8}$ A & -4.639 & 24.49 & unblended\\
         CH$_3$C(O)NH$_2$ & 97.943874 & $9_{1,9}-$8$_{1,8}$ A & -5.846 & 24.49 & unblended\\
         CH$_3$C(O)NH$_2$ & 97.943882 & $9_{0,9}-$8$_{0,8}$ A & -5.846 & 24.49 & unblended\\
         CH$_3$C(O)NH$_2$ & 97.943883 & $9_{1,9}-$8$_{0,8}$ A & -4.639 & 24.49 & unblended\\
         CH$_3$C(O)NH$_2$ & 108.214105 & $9_{1,8}-$8$_{2,7}$ A & -4.625 & 28.66 & blended with C$_2$H$_5$CN\\
         CH$_3$C(O)NH$_2$ & 108.214251 & $9_{2,8}-$8$_{2,7}$ A & -5.803 & 28.66 & blended with C$_2$H$_5$CN\\
         CH$_3$C(O)NH$_2$ & 108.215083 & $9_{1,8}-$8$_{1,7}$ A & -5.803 & 28.66 & blended with C$_2$H$_5$CN\\
         CH$_3$C(O)NH$_2$ & 108.215230 & $9_{2,8}-$8$_{1,7}$ A & -4.625 & 28.66 & blended with C$_2$H$_5$CN\\
         CH$_3$C(O)NH$_2$ & 108.255232 & $10_{0,10}-$9$_{1,9}$ A & -4.51 & 29.68 & unblended\\
         CH$_3$C(O)NH$_2$ & 108.255232 & $10_{1,10}-$9$_{1,9}$ A & -5.719 & 29.68 & unblended\\
         CH$_3$C(O)NH$_2$ & 108.255234 & $10_{0,10}-$9$_{0,9}$ A & -5.719 & 29.68 & unblended\\
         CH$_3$C(O)NH$_2$ & 108.255234 & $10_{1,10}-$9$_{0,9}$ A & -4.510 & 29.68 & unblended\\
         CH$_3$C(O)NH$_2$ & 110.952601 & $5_{5,1}-$4$_{4,0}$ A & -4.828 & 15.69 & blended with CH$_3$COOH\\
         CH$_3$C(O)NH$_2$ & 113.868958 & $7_{4,4}-$6$_{3,3}$ A & -4.972 & 23.61 & share with  CH$_3$C(O)NH$_2$ $5_{5,1}-$4$_{4,1}$ E \\
         CH$_3$C(O)NH$_2$ & 139.138720 & $12_{1,11}-$11$_{2,10}$ A & -4.281 & 47.21 & blended with HCCCH$_2$CN\\
         CH$_3$C(O)NH$_2$ & 139.138723 & $12_{2,11}-$11$_{1,10}$ A & -4.281 & 47.21 & blended with HCCCH$_2$CN\\
         CH$_3$C(O)NH$_2$ & 139.188418 & $13_{0,13}-$12$_{1,12}$ A & -4.199 & 48.24 & unblended\\
         CH$_3$C(O)NH$_2$ & 139.188418 & $13_{1,13}-$12$_{0,12}$ A & -4.199 & 48.24 & unblended\\
         \hline
        trans-N-CH$_3$NHCHO & 32.550890 & 3$_{1,3}-$2$_{1,2}$ E & -6.223 & 7.48 & unblended\\
        trans-N-CH$_3$NHCHO & 33.455906 & 3$_{0,3}-$2$_{0,2}$ A & -6.154 & 3.23 & unblended\\
        trans-N-CH$_3$NHCHO & 34.461458 & 3$_{1,2}-$2$_{1,1}$ E & -6.230 & 8.42 & unblended\\
        trans-N-CH$_3$NHCHO & 34.693486 & 2$_{1,2}-$1$_{0,1}$ A & -6.537 & 2.21 & unblended\\
        trans-N-CH$_3$NHCHO & 37.428359 & 4$_{2,2}-$4$_{1,3}$ A & -6.200 & 8.26 & blended with C$_6$H\\
        trans-N-CH$_3$NHCHO & 39.138171 & 3$_{2,1}-$3$_{1,2}$ A & -6.343 & 6.04 & blended with U\\
        trans-N-CH$_3$NHCHO & 43.793726 & 3$_{1,3}-$2$_{0,2}$ A & -6.199 & 3.72 & unblended\\
        trans-N-CH$_3$NHCHO & 44.092441 & 4$_{0,4}-$3$_{0,3}$ A & -5.793 & 5.34 & unblended\\
        trans-N-CH$_3$NHCHO & 45.131824 & 4$_{2,3}-$3$_{2,2}$ A & -5.899 & 8.18 & unblended\\
        trans-N-CH$_3$NHCHO & 46.120191 & 5$_{0,5}-$4$_{1,4}$ A & -3.990 & 7.95 & blended with HOCH$_2$CN\\
        trans-N-CH$_3$NHCHO & 72.670325 & 7$_{1,7}-$6$_{1,6}$ A & -5.135 & 14.74 & unblended\\
        trans-N-CH$_3$NHCHO & 74.090940 & 7$_{0,7}-$6$_{0,6}$ A & -5.115 & 14.60 & unblended\\
        trans-N-CH$_3$NHCHO & 74.122946 & 7$_{0,7}-$6$_{0,6}$ E & -5.094  & 18.24 & blended with $^{13}$CH$_2$CHCN\\
        trans-N-CH$_3$NHCHO & 79.741415 & 3$_{2,1}-$2$_{1,2}$ A & -5.847 & 6.03 & unblended\\
        trans-N-CH$_3$NHCHO & 82.715063 & 8$_{1,8}-$7$_{1,7}$ A & -4.969 & 18.71 & blended with H$_2$CCCHCN\\
        trans-N-CH$_3$NHCHO & 83.253887 & 7$_{2,5}-$6$_{2,4}$ A & -5.047 & 18.45 & unblended\\
        trans-N-CH$_3$NHCHO & 87.496110 & 8$_{1,8}-$7$_{0,7}$ E & -5.204 & 22.44 & unblended\\
        trans-N-CH$_3$NHCHO & 89.031261 & 8$_{2,7}-$7$_{2,6}$ A & -4.930 & 22.13 & unblended\\
        trans-N-CH$_3$NHCHO & 89.935665 & 8$_{1,7}-$7$_{1,6}$ E & -4.886 & 24.61 & unblended\\
        trans-N-CH$_3$NHCHO & 92.415560 & 9$_{1,9}-$8$_{1,8}$ E & -4.817 & 26.87 & unblended\\
        trans-N-CH$_3$NHCHO & 93.395965 & 8$_{1,7}-$7$_{1,6}$ A & -4.869 & 21.18 & unblended\\
        trans-N-CH$_3$NHCHO & 93.406380 & 9$_{0,9}-$8$_{0,8}$ A & -4.816 & 23.10 & unblended\\
        trans-N-CH$_3$NHCHO & 95.642981 & 8$_{2,6}-$7$_{2,5}$ A & -4.866 & 23.04 & blended with C$_2$H$_5$OCH$_3$\\
        trans-N-CH$_3$NHCHO & 105.608314 & 3$_{3,0}-$2$_{2,1}$ A & -5.362 & 9.45 & unblended\\
        trans-N-CH$_3$NHCHO & 107.852547 & 9$_{2,7}-$8$_{2,6}$ A & -4.712 & 28.22 & unblended\\
        trans-N-CH$_3$NHCHO & 109.325438 & 3$_{3,0}-$2$_{2,0}$ E & -5.314 & 14.35 & unblended\\
         \hline
         \hline
         NH$_2^{13}$CHO & 40.778439 & 2$_{1,2}-$1$_{1,1}$ & -4.664 & 5.8 & unblended\\
         NH$_2^{13}$CHO & 42.318061 & 2$_{0,2}-$1$_{0,1}$ & -4.503 & 3.0 & unblended\\
         NH$_2^{13}$CHO & 81.495632 & 4$_{1,4}-$3$_{1,3}$ & -3.674 & 12.7 & unblended\\
         NH$_2^{13}$CHO & 84.390679 & 4$_{0,4}-$3$_{0,3}$ & -3.612 & 10.1 & unblended\\
         NH$_2^{13}$CHO & 101.813583 & 5$_{1,5}-$4$_{1,4}$ & -3.379 & 17.6 & unblended\\
         NH$_2^{13}$CHO & 105.260270 & 5$_{0,5}-$4$_{0,4}$ & -3.330 & 15.2 & unblended\\
         \hline
         NH$_2$C(O)NH$_2$ & 37.926700 & 3$_{0,3}-$2$_{0,2}$ & -5.011 & 3.89 & unblended\\
         NH$_2$C(O)NH$_2$ & 39.116400 & 2$_{2,1}-$1$_{1,0}$ & -5.195 & 2.91 & blended with U\\
         NH$_2$C(O)NH$_2$ & 48.697422 & 4$_{0,4}-$3$_{1,3}$ & -4.648 & 6.23 & unblended\\
         NH$_2$C(O)NH$_2$ & 48.705728 & 4$_{1,4}-$3$_{0,3}$ & -4.648 & 6.23 & blended with U\\
         $^*$NH$_2$C(O)NH$_2$ & 81.104130 & 6$_{1,5}-$5$_{2,4}$ & -4.107 & 15.29 & blended with DNCO\\
         $^*$NH$_2$C(O)NH$_2$ & 81.108770 & 6$_{2,5}-$5$_{1,4}$ & -4.107 & 15.29 & blended with DNCO\\
         $^*$NH$_2$C(O)NH$_2$ & 81.199200 & 7$_{0,7}-$6$_{1,6}$ & -3.945 & 16.36 & unblended\\
         $^*$NH$_2$C(O)NH$_2$ & 81.199200 & 7$_{1,7}-$6$_{0,6}$ & -3.945 & 16.36 & unblended\\
         NH$_2$C(O)NH$_2$ & 92.031820 & 8$_{0,8}-$7$_{1,7}$ & -3.780 & 20.78 & blended with CH$_2$CH$^{13}$CN\\
         NH$_2$C(O)NH$_2$ & 92.031820 & 8$_{1,8}-$7$_{0,7}$ & -3.780 & 20.78 & blended with CH$_2$CH$^{13}$CN\\
         NH$_2$C(O)NH$_2$ & 96.185230 & 5$_{4,2}-$4$_{3,1}$ & -4.253 & 14.53 & blended with U\\
         $^*$NH$_2$C(O)NH$_2$ & 102.76756 & 8$_{1,7}-$7$_{2,6}$ & -3.754 & 24.64 & blended with CH$_3$COCH$_3$\\
         $^*$NH$_2$C(O)NH$_2$ & 102.76756 & 8$_{2,7}-$7$_{1,6}$ & -3.754 & 24.64 & blended with CH$_3$COCH$_3$\\
         $^*$NH$_2$C(O)NH$_2$ & 102.86432 & 9$_{0,9}-$8$_{1,8}$ & -3.635 & 25.71 & unblended\\
         $^*$NH$_2$C(O)NH$_2$ & 102.86432 & 9$_{1,9}-$8$_{0,8}$ & -3.635 & 25.71 & unblended\\
    \label{tab:mol_trans}
   \end{longtable}
%


\bsp	
\label{lastpage}
\end{document}